\documentclass{aa}  

\usepackage{graphicx}
\usepackage{txfonts}
\usepackage{lipsum}
\usepackage{subcaption}         
\usepackage{lscape}             
\usepackage{placeins}           

\usepackage[colorlinks=true,citecolor=blue]{hyperref}
\usepackage{siunitx}
\usepackage{xcolor}
\usepackage{orcidlink}
\usepackage{natbib}
\bibpunct{(}{)}{;}{a}{}{,} 

\newcommand\grizli{\ensuremath{\textsc{grizli}}}
\newcommand\bagpipes{\ensuremath{\textsc{bagpipes}}}

\newcommand\zorig{\ensuremath{z_{\rm{orig}}}}

\newcommand\zspec{\ensuremath{z_{\rm{spec}}}}
\newcommand\zniriss{\ensuremath{z_{\rm{NIRISS}}}}

\newcommand{\Oiii}{\ensuremath{\left[\ion{O}{iii}\right]}}
\newcommand{\Oii}{\ensuremath{\left[\ion{O}{ii}\right]}}

\newcommand{\Oiiiblue}{\ensuremath{\left[\ion{O}{iii}\right]\,{\lambda 4959}}}
\newcommand{\Oiiired}{\ensuremath{\left[\ion{O}{iii}\right]\,{\lambda 5007}}}
\newcommand{\Niiblue}{\ensuremath{\left[\ion{N}{ii}\right]\,{\lambda 6548}}}
\newcommand{\Niired}{\ensuremath{\left[\ion{N}{ii}\right]\,{\lambda 6584}}}
\newcommand{\Nii}{\ensuremath{\left[\ion{N}{ii}\right]}}
\newcommand{\Sii}{\ensuremath{\left[\ion{S}{ii}\right]}}
\newcommand{\Siii}{\ensuremath{\left[\ion{S}{iii}\right]}}
\newcommand{\Neiii}{\ensuremath{\left[\ion{Ne}{iii}\right]}}
\newcommand{\Halpha}{\ensuremath{\ion{H}{$\alpha$}}}
\newcommand{\Hbeta}{\ensuremath{\ion{H}{$\beta$}}}

\newcommand{\Hii}{\ensuremath{\ion{H}{ii}}}

\newcommand{\sn}{\ensuremath{\mathrm{S/N}}}
\newcommand{\chisq}{\ensuremath{\chi^2}}
\newcommand{\redchisq}{\ensuremath{\chi^2_{\nu}}}
\newcommand{\niters}{\ensuremath{n_{\rm{iters}}}}
\newcommand{\nsamples}{\ensuremath{n_{\rm{samples}}}}
\newcommand{\nregions}{\ensuremath{n_{\rm{regions}}}}
\newcommand{\BD}{\ensuremath{\rm{BD}}}
\newcommand{\re}{\ensuremath{r_{\rm{e}}}}
\newcommand{\edens}{\ensuremath{n_{\rm{e}}}}
\newcommand{\EBV}{\ensuremath{E(B-V)}}
\newcommand{\EBVneb}{\ensuremath{E(B-V)_{\mathrm{neb}}}}
\newcommand{\EBVsed}{\ensuremath{E(B-V)_{\mathrm{SED}}}}
\newcommand{\klam}{\ensuremath{\kappa\left(\lambda\right)}}
\DeclareRobustCommand{\klam}[1]{\ensuremath{\kappa\left(\lambda_{#1}\right)}}
\newcommand\logstellmass{\ensuremath{\log_{10}\left(M_*/\Msol\right)}}
\newcommand\sfrunits{\ensuremath{\Msol\,\mathrm{yr}^{-1}}}
\newcommand\Msol{\ensuremath{\mathrm{M}_{\sun}}}
\newcommand\Zsol{\ensuremath{\mathrm{Z}_{\sun}}}
\newcommand\sfrsurfdens{\ensuremath{\Sigma_{\mathrm{SFR}}}}
\newcommand\logsfrsurfdens{\ensuremath{\log_{10}\Sigma_{\mathrm{SFR}}}}
\newcommand\logmstarsurfdens{\ensuremath{\log_{10}\Sigma_*}}

\defcitealias{watson_glass-jwst_2025}{W25a}

\begin{document}

    \title{Spatially Resolved Nebular-Stellar Reddening with JWST/NIRISS}


\author{
    Peter J. Watson    \inst{1}\corrauth{peter.watson@inaf.it}\orcidlink{0000-0003-3108-0624}
    \and
    Benedetta Vulcani
    \inst{1}\email{benedetta.vulcani@inaf.it}\orcidlink{0000-0003-0980-1499}
    \and
    Tommaso Treu
    \inst{2}\email{tt@astro.ucla.edu }\orcidlink{0000-0002-8460-0390}
    \and
    Ayan Acharyya
    \inst{1}\email{ayan.acharyya@inaf.it}\orcidlink{0000-0003-4804-7142}
    \and
    Marc Rafelski
    \inst{3,4}\email{mrafelski@stsci.edu}\orcidlink{0000-0002-9946-4731}    
    \and
    Anahita Alavi
    \inst{5}\email{anahita@ipac.caltech.edu}\orcidlink{0000-0002-8630-6435}
    \and
    Matthew Hayes
    \inst{6}\email{matthew.hayes@astro.su.se}\orcidlink{0000-0001-8587-218X}
    \and
    Keunho Kim
    \inst{5}\email{keunho11@ipac.caltech.edu}\orcidlink{0000-0001-6505-0293}
    \and
    Faezeh Manesh
    \inst{7}\email{faezehsadat.akhlaghimanesh@email.ucr.edu}\orcidlink{0009-0008-4976-3216}
    \and
    Claudia Scarlata
    \inst{8}\email{mscarlat@umn.edu}\orcidlink{0000-0002-9136-8876}
}

\institute{
    INAF -- Osservatorio Astronomico di Padova, Vicolo Osservatorio 5, 35122 Padova, Italy
    \and  
    University of California, Los Angeles, Department of Physics and Astronomy, 430 Portola Plaza, Los Angeles, CA 90095, USA
    \and
    Space Telescope Science Institute, 3700 San Martin Drive, Baltimore, MD 21218, USA
    \and 
    Department of Physics and Astronomy, Johns Hopkins University, Baltimore, MD 21218, USA
    \and
    IPAC, Mail Code 314-6, California Institute of Technology, 1200 E. California Blvd., Pasadena, CA 91125, USA
    \and
    Stockholm University, Department of Astronomy, AlbaNova University Center, SE-106 91 Stockholm, Sweden
    \and
    Department of Physics and Astronomy, University of California, 900 University Ave, Riverside, CA 92521, USA
    \and
    Minnesota Institute for Astrophysics, School of Physics and Astronomy, University of Minnesota, 316 Church Street SE, Minneapolis, MN 55455, USA
}

    \date{Received 10 June 2026 / Accepted 13 September 2026}

    \abstract{
        An accurate determination of the dust attenuation within galaxies is essential to derive key physical properties such as the star formation rate (SFR).
        We present an analysis using the JWST/NIRISS data from the GLASS-JWST ERS programme to investigate and characterise the stellar and nebular reddening of galaxies at $1.0<z<2.4$, down to the sub-kpc scale.
        We use a multiregion fitting method to extract high-quality \Halpha\ and \Hbeta\ emission line maps for 99 individual galaxies across a stellar mass range $7.0<\logstellmass<10.5$.
        We find no evidence for ratios of the Balmer decrement (\Halpha/\Hbeta) below the intrinsic limit for Case B recombination, beyond the expected variation from observational uncertainties.
        We reproduce the local correlation between the Balmer decrement and total stellar mass, 
        and find no measurable difference when splitting the sample by redshift, with negligible attenuation below $\logstellmass\lesssim8.5$.
        Similarly, the best-fit relation between the nebular and continuum reddening follows the same relation as in local starburst galaxies, $\EBVsed=(0.46\pm0.02)\EBVneb$, together indicating no significant evolution in the dust geometry within galaxies out to $z\lesssim2.4$.
        We derive best-fit linear relations between the differential nebular-stellar reddening and the SED-derived star formation rate (SFR) and stellar mass, finding statistically significant relations for both quantities.
        We use our spatially-resolved measurements to derive an empirical calibration between the resolved 
        differential reddening, and the 
        SFR surface density.
        These will enable crucial dust attenuation corrections for spatially-resolved science at higher redshifts where the Balmer lines are inaccessible, such as with future Roman grism observations.
    }


    \maketitle
    \nolinenumbers

\section{Introduction} \label{sec:intro}
Understanding when and where star formation occurs is crucial to understanding the growth and evolution of galaxies over cosmic time. 
This is often traced using star formation rate (SFR) indicators such as the Hydrogen recombination lines or the UV continuum flux,
which together underpin many of our observational constraints on the star formation history (SFH) of the universe.
The Hydrogen recombination lines arise from the photoionised gas of \Hii\ regions, predominantly from short-lived stars of $\gtrsim15\,\Msol$, and hence tracking SF on timescales $\lesssim10$Myr \citep{chabrier_galactic_2003,kennicutt_star_2012}.
The Balmer lines, transitions to the $n=2$ level of Hydrogen, are favourably located in the optical spectrum, and therefore can be used to measure SFRs out to $z$\,$\gtrsim$\,6 \citep{sun_first_2023,pirie_jwst_2025}.
However, these measurements suffer from systematic uncertainties, most noticeably attenuation from dust.
Dust grains in and around the \Hii\ regions surrounding young stars absorb and scatter the nebular emission, preventing direct measurements of the total recombination flux. 
This attenuation varies strongly with wavelength, and depends on both the size distribution of dust grains and their spatial distribution with respect to the emitting source \citep{lorenz_measuring_2025,reddy_aurora_2026a}.

If multiple recombination lines are observed, it is possible to estimate the shape of the wavelength-dependent nebular attenuation curve \citep{charlot_simple_2000,calzetti_dust_2000,reddy_mosdef_2015,battisti_average_2022,reddy_aurora_2026a,rodighiero_first_2026},
by comparing against the intrinsic dust-free ratios.
Although the Balmer line ratio of \Halpha\ to \Hbeta, also known as the Balmer decrement, is the most common, the higher-order IR recombination series such as the Paschen and Brackett lines have seen increasing use since the launch of JWST \citep[e.g.][]{reddy_paschen-line_2023,lin_nebular_2024,seille_physical_2024, liu_characterizing_2026,reddy_aurora_2026b}.
The intrinsic values of these line ratios are typically calculated under the assumption that the nebula is optically thick to the Lyman series (transitions to $n=1$), known as the Case B recombination limit \citep{menzel_physical_1937,baker_physical_1938}.
Although other recombination theories exist, these are very rarely observed in optical emission lines \citep[e.g.][]{zhao_radio_1996}.
Case A, for example, in which the nebula is optically thin to Lyman transitions, leads to substantially lower fluxes in the Balmer series and is more applicable to diffuse gas outside \Hii\ regions (e.g. \citealt{patel_are_1995}; see also \citealt{scarlata_universal_2024} for a thorough discussion of alternatives to Case B recombination).
It has long been taken for granted that Case B recombination accurately describes the overall Balmer line emission from galaxies, but 
this is not always the case: 
as recent space-based surveys across a range of redshifts have reported up to 40\% of line ratios inconsistent with Case B \citep{pirzkal_next_2024, shapley_jwstnirspec_2023,sun_mysteriously_2025,sandles_jades_2024,woodrum_jades_2025,llerena_extreme_2026}.
If this fundamental assumption is not universally valid, then dust attenuation corrections may be incorrect, compromising our ability to measure the intrinsic star formation of galaxies.

Recombination line ratios are not the only measure of attenuation, however.
Attenuation of the stellar continuum can be readily inferred from broad-band spectral energy distribution (SED) fitting \citep[e.g.][]{kriek_dust_2013,carnall_inferring_2018,abdurrouf_dissecting_2022,werle_history_2024}.
Comparing the stellar to nebular attenuation can provide crucial constraints on the distribution of dust within a galaxy and around star-forming regions \citep{price_direct_2014,battisti_average_2022}.
In local star-forming galaxies, it is well-established that the nebular emission is typically more strongly attenuated than the stellar component \citep{calzetti_dust_2000,groves_balmer_2012,lin_variant_2020}.
This is often interpreted through a two-component dust model \citep{charlot_simple_2000}, in which the interstellar medium (ISM) is composed of both a diffuse component, and dense molecular birth clouds.
Older stars are attenuated only by the diffuse ISM, whereas \Hii\ regions embedded within birth clouds are attenuated by both components, naturally leading to a higher nebular-to-stellar attenuation. 
Some analyses have indicated that this relationship, and therefore the spatial distribution of dust within galaxies, shows little evolution up to $z\sim2$ \citep{dominguez_dust_2013,price_direct_2014,shapley_mosfire_2022,alavi_uv_2026}, or even beyond to $z\sim7$ \citep{shapley_jwstnirspec_2023,sandles_jades_2024,woodrum_jades_2025,karthikeyan_balmer_2026}.

However, these integrated measurements can mask complex internal variations. 
Consequently,
exploring the relationship between nebular and stellar attenuation in a spatially-resolved manner is vital to understanding the internal distribution of dust within galaxies and how it evolves over cosmic time.
One increasingly useful tool for this is wide-field slitless spectroscopy (WFSS), an invaluable observing mode on space-based telescopes (e.g. HST, JWST, Euclid, and Roman).
In this mode, the light from an entire field of view is simultaneously dispersed onto a detector, enabling the efficient collection of large, spatially-resolved spectroscopic samples.
However, at Cosmic Noon ($1\lesssim z\lesssim3$), the peak of SF in the universe, the wavelength coverage of many of these instruments frequently prevents the observation of multiple Balmer lines at a reliable signal-to-noise (\sn), and thus a direct derivation of the nebular attenuation.
Obtaining empirical calibrations for the nebular reddening at these redshifts is therefore essential to unlock the full potential of these instruments.

To address this need, in this work, we use data from the GLASS-JWST ERS programme \citep{treu_glass-jwst_2022}, specifically the WFSS observations using the Near Infrared Imager and Slitless Spectrograph \citep[NIRISS,][]{doyon_jwst_2012}, described in more detail in \cite{roberts-borsani_early_2022,boyett_early_2022,watson_glass-jwst_2025}.
We use these data to investigate both adherence to Case B recombination theory, and derive new calibrations for the nebular-stellar attenuation.
This paper is organised as follows.
In Section~\ref{method}, we describe the primary and ancillary data, as well as the sample selection, and the reduction process and methods used to obtain the final data products.
In Section~\ref{results}, we compare the line fluxes obtained via different methods, evaluate the consistency of our sample with Case B recombination theory, and derive new empirical calibrations.
We discuss the implications and scientific context of these in Section~\ref{discuss}, and summarise the analysis in Section~\ref{conclusions}.

Throughout this analysis, coordinates are given in the International Celestial Reference System (ICRS); magnitudes are in the AB system; and we assume a standard $\Lambda$ cold dark matter cosmology, with $\Omega_{\rm{M}}=0.3$, $\Omega_{\Lambda}=0.7$, and $h=0.7$.

\section{Method} \label{method}

\subsection{Data reduction} \label{method:data_reduction}

The NIRISS data reduction follows a similar method to that described in \citet[\citetalias{watson_glass-jwst_2025} hereinafter]{watson_glass-jwst_2025}, and we elaborate only on the updates here.
The data were reduced from uncalibrated ramp exposures to count-rate files using the \textsc{Detector1} module from the \textsc{jwst} package\footnote{\href{https://github.com/spacetelescope/jwst}{\texttt{github.com/spacetelescope/jwst}}, version 1.20.2}, with the operational context ``1467.pmap''.
Further processing of the data was performed using a modified version of the Grism Redshift \& Line package\footnote{\href{https://github.com/PJ-Watson/grizli}{\texttt{github.com/PJ-Watson/grizli}}, version 1.13.3} \citep[\grizli;][]{brammer_grizli_2019}, based on version 1.13.3 of the original.
The trace location and wavelength calibration used a combination of the NGDEEP configuration files \citep{pirzkal_next_2024} for the 1st order dispersed spectra, and the configuration files from \cite{matharu_updated_2022} for all other spectral orders.
Extensive testing found that this combination gave the best compromise between a reliable modelling of the contamination from the 0th spectral order, and an accurate wavelength calibration for the 1st order, consistent across both grisms and all blocking filters.
We examine in further detail the differences between several versions of the reduced data in Appendix~\ref{app:pipeline_differences}. 

To account for the contamination from the dispersed intracluster light, we began by adopting the same segmentation map as used in \citetalias{watson_glass-jwst_2025}, which also allowed for consistent source IDs between the two analyses.
We then subtracted a diffuse background from each direct image using \textsc{Background2D} from the \textsc{photutils} package, with a maximum threshold of $5\sigma$ above the sky background to avoid oversubtraction of the extended cluster galaxies.
This background was forward modelled using \grizli, assuming a flat $f_{\lambda}$ spectrum, and subtracted from all grism observations.
The full contamination model was then forward-modelled for all sources, similarly adopting a flat $f_{\lambda}$ spectrum.
Although fitting a polynomial model spectrum for all sources should in theory give a more accurate contamination model, we found this to be out-performed by a flat spectrum, as this reduced fitting artefacts associated with saturated sources (such as foreground stars or AGN) in the direct imaging.
Sources of interest were then forward-modelled and extracted using \grizli\ following an identical procedure as that described in \cite{roberts-borsani_early_2022} and \citetalias{watson_glass-jwst_2025}, where pixels used for extraction were weighted by
\begin{equation}
    w_i = \sigma_i^{-2}\cdot\exp\left(-a\cdot|f_{\mathrm{cont,}i}|/\sigma_i\right),
\end{equation}
for pixels $i$ with flux uncertainty $\sigma_i$, contaminating flux $f_{\mathrm{cont,}i}$, and a constant factor $a=0.2$.
As in \citetalias{watson_glass-jwst_2025}, we did not rescale the extracted spectra to match the broadband photometry, as the derived correction factors for these data were negligible.
This procedure produced 2D dispersed spectra, emission-line maps, and best-fit integrated line fluxes, which we refer to hereinafter as the ``\grizli'' sample.

\subsection{Ancillary data} \label{method:ancillary_data}

We also make use of the publicly available drizzled mosaics from the MegaScience Data Release \citep{suess_medium_2024}, for broad-band SED fitting.
In order to avoid any mismatch in spatial resolution between the imaging and the NIRISS/WFSS data, we use only filters where the point-spread function (PSF) is narrower than that of the longest-wavelength NIRISS filter (F200W).
As such, we use the cluster galaxy-subtracted mosaics in the HST/ACS bands F435W, F606W, and F814W, and JWST/NIRCam F070W, F090W, F115W, F140M, F150W, F162M, F182M, and F200W.
These maps were originally reduced using the ``jwst\_0995.pmap'' operational context, and aligned and co-added using \grizli\ onto a grid with a scale of 0\farcs 04/pixel.
We examine this choice of filters in Appendix~\ref{app:filter_selection}, and conclude that the absence of the longer-wavelength data has no significant impact on our resulting analysis.
We used \textsc{pypher} \citep{boucaud_convolution_2016} to derive convolution kernels matching the PSF of each filter to that of JWST/NIRISS F200W, and \textsc{astropy} and \textsc{reproject} to convolve and align the mosaics to the same pixel grid as that used in our reduction of the NIRISS data.

\subsection{Sample selection} \label{method:sample_selection}

\begin{figure}
    \centering
    \includegraphics[width=\columnwidth]{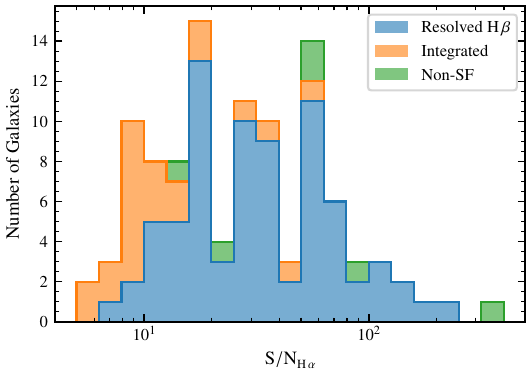}
    \caption{
    The \sn\ distribution of the integrated \Halpha\ fluxes, as measured by \grizli, shown as a stacked histogram.
    We separate out galaxies that were determined to have non-SF ionisation (green), galaxies for which integrated \Hbeta\ fluxes could be measured (both blue and orange), and galaxies where \Hbeta\ could be resolved above the background level (blue, see Section~\ref{results:case_b:resolved}).
    }
    \label{fig:sample_selection_SNR}
\end{figure}

For our initial sample selection, we consider only galaxies which were determined to have a secure spectroscopic redshift in the catalogue by \citetalias{watson_glass-jwst_2025}.
However, we first merged several source IDs in this existing catalogue, listed in Appendix~\ref{app:interacting_systems}.
These were sources at identical redshifts ($\delta z/(1+z)<1\times10^{-3}$), small on-sky separations ($<1\arcsec$), and for which it was not possible to reliably extract individual dispersed spectra without strong residual contamination from the neighbouring object(s).
The exact physical nature of these sources, whether they are interacting systems or components of a single clumpy galaxy, is not relevant to this analysis, but will be explored in a separate paper (E. Golden-Marx et al., \textit{in prep.}).

We require that \Halpha\ is detected with an integrated $\sn>5$ (using the default 2D \grizli\ fits), and do not place any initial constraints on the \Hbeta\ \sn\ to avoid biasing our sample towards galaxies with a lower integrated dust attenuation.
We also consider the difficulty in both obtaining reliable flux measurements in regions with low filter throughput, and the persistent uncertainties in the NIRISS wavelength calibration (\citealt{noirot_first_2023, pirzkal_next_2024}; \citetalias{watson_glass-jwst_2025}; \citealt{acharyya_spatially_2026, malkan_parallel_2025}; also Appendix~\ref{app:pipeline_differences}).
As such, we reject galaxies where the centroid of either line (\Halpha\ or \Hbeta) lies in the wavelength range where the filter sensitivity drops to less than 50\% of the peak value.
These initial selection criteria are satisfied by 105 sources, including the triply-imaged galaxy in \citetalias{watson_glass-jwst_2025}, with the distribution of the integrated \Halpha\ \sn\ shown in Fig.~\ref{fig:sample_selection_SNR}.
The median \sn\ of \Halpha\ in this sample is 27, and for \Hbeta, $\sn=6.4$.

\begin{figure}
    \centering
    \includegraphics[width=\columnwidth]{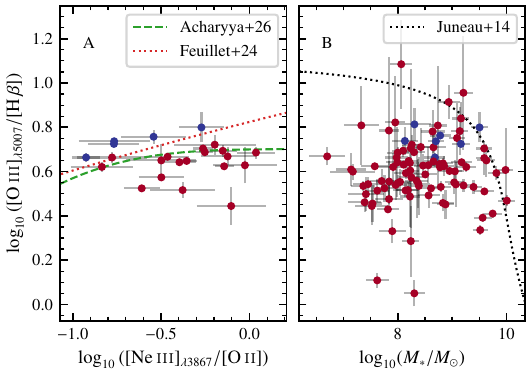}
    \caption{
    (A) The distribution of galaxies in the \textit{OHNO} parameter space, using the \grizli-measured line fluxes.
    Galaxies are only shown if the measured fluxes have a $\sn>=2$ for all four lines. 
    The dashed green line and dotted red lines indicate the AGN-SF demarcation of \cite{acharyya_spatially_2026} and \cite{feuillet_classifying_2024} respectively.
    Sources lying more than 1$\sigma$ above the former were considered to be inconsistent with ionisation purely from star formation, and were discarded from the sample.
    (B) The distribution of galaxies in the mass-excitation (MEx) plane, with the upper limit on SF galaxies from \cite{juneau_active_2014} plotted as a dotted line.
    Galaxies plotted in blue are those that are rejected by the \textit{OHNO} demarcation.
    }
    \label{fig:ohno_sample}
\end{figure}

From this selection, we remove sources where the nebular emission is inconsistent with ionisation from star formation, such as from AGN.
The only previously known AGN in this sample is ID 321, associated with an X-ray point source (see the discussion in \citetalias{watson_glass-jwst_2025}).
Whilst the diagnostic diagrams of \cite{baldwin_classification_1981} are frequently employed for this purpose,
at the NIRISS spectral resolution of $R\sim150$, we cannot resolve the required lines
(see Section~\ref{method:line_flux_corrections}).
Instead, we employ the \textit{OHNO} demarcation introduced in \cite{acharyya_spatially_2026}, which denotes the upper limit of the MAPPINGS v5.1 \Hii\ region photoionisation models in the \Oiii/\Hbeta-\Neiii/\Oii\ parameter space.
Due to the redshift range covered by our sample, these lines are only available at a $\sn>=2$ (using the integrated \grizli\ fluxes) for 23 sources, shown in Fig.~\ref{fig:ohno_sample}A.
Five of the sources in this sample (IDs 1156, 1721, 1814, 2938, and 3495) lie more than 1$\sigma$ above the AGN-SF demarcation, and were therefore removed from our analysis.
This is consistent with the \textit{OHNO} demarcation independently derived by \cite{feuillet_classifying_2024} using SDSS data, which also rejects four of the five sources as shown in Fig.~\ref{fig:ohno_sample}A.

As this AGN selection provides only a lower limit on the number of sources in our sample with non-SF ionisation, we also test against the mass-excitation (MEx) diagram of \cite{juneau_active_2014}, as shown in Fig.~\ref{fig:ohno_sample}B.
This diagnostic replaces the \Nii/\Halpha\ ratio used by \cite{baldwin_classification_1981} with stellar mass, producing an AGN/SF demarcation in the \Oiii/\Hbeta-\logstellmass\ space, calibrated up to $z\sim2$.
We find an unusually poor correspondence between the diagnostics, with all galaxies in the MEx parameter space lying in the SF region within uncertainties, including those rejected by the \textit{OHNO} diagnostics.
We defer speculation as to the reasons behind this for future analyses, noting that this discrepancy between diagnostics has also been observed in other samples at $z\gtrsim1$ \citep[][K. Nedkova et al., \textit{in prep.}]{llerena_extreme_2026}.
We do, however, perform a visual inspection of the five rejected sources, and confirm that the 2D spectral fits and resulting emission line maps show clear signs of template mismatch, as expected in the case where emission from non-SF ionisation dominates \citep{noirot_first_2023}.
We therefore perform the multiregion fitting (see Section~\ref{method:multiregion}) for the remaining 99 sources.
As we discuss later in Section~\ref{results:case_b:resolved}, this is reduced to 76 sources for the spatially-resolved portions of our analysis, as we are not able to construct \Hbeta\ emission line maps for 27 sources where this is detectable above the background level.

\subsection{Multi-region fitting} \label{method:multiregion}

\begin{figure*}
    \centering
    \includegraphics[trim={1in 0.7in 1in 1in},clip,width=\textwidth]{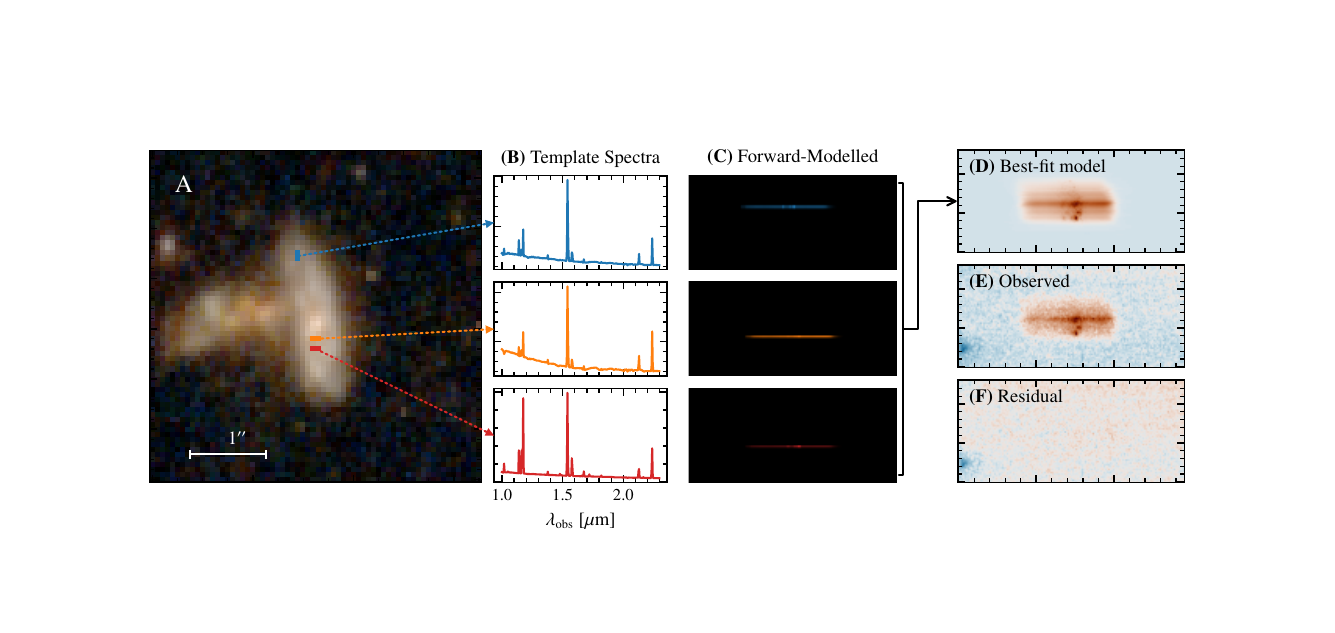}
    \caption{
    {An illustration of the multi-region approach to fitting NIRISS/WFSS data.
        (A) An RGB image of one galaxy (ID 3070), showing the locations of three example regions used for the 2D SED fitting. The full segmentation map for this galaxy consists of 185 such regions.
        (B) Sample template spectra generated from the SED fitting posteriors, covering the full wavelength range probed by NIRISS.
        (C) The forward-modelled grism spectra, in a single filter (F115W; $\approx\,$1--1.3\textmu m) and grism orientation (GR150R).
        (D) The optimum combination of all forward-modelled templates, compared against the observed grism spectrum (E), and the residual (observed$-$model) spectrum (F), all of which are plotted using the same colour map.
    }
    }
    \label{fig:forward_modelling_example}
\end{figure*}

In the standard procedure for spectral extraction of WFSS data, as implemented in \grizli\ and the STScI pipelines, one of the principal assumptions is that the shape of the SED is uniform across the extent of the source, and hence that all nebular and stellar emission is co-located.
When working with low \sn\ data, or sources not substantially larger than the PSF, this enables an efficient extraction of 1D spectra, or 2D (continuum-subtracted) emission line maps.
However, it is evidently a poor assumption for any significantly extended source in which the spatial variation of the spectrum can be resolved, and this limitation has been discussed thoroughly in a number of previous studies \citep[e.g.][]{sorba_using_2017,pirzkal_two-dimensional_2018,bagley_hst_2020}.
Most recently, \cite{estrada-carpenter_when_2024} introduced a method whereby a full spatially-resolved fit can be obtained, by a two-stage process involving a resolved SED fit to construct sets of template spectra, and an iterative forward-modelling procedure based around \grizli.

\begin{table}[]
    \centering
    \begin{tabular}{c c m{2.5cm}}
    Parameter & Value & Prior \\\hline\hline
    Redshift & $\zspec\pm2.5\times10^{-3}$ & Uniform \\ 
    \hline
    \multicolumn{3}{c}{Continuity Star Formation History}\\
    \hline
    \logstellmass\ & (5, 11) & Uniform \\
    $Z/\Zsol$ & (0.0, 3.0) & Gaussian \mbox{($\mu=1.0$, $\sigma=0.5$)} \\
    Age Bins & 5 & Logarithmic \\
    Age of 1st Bin & 20\,Myr & Fixed \\
    $\Delta\log\mathrm{SFR}$ & (-10,10)& Student's-$t$ \mbox{($\nu=2$, $\sigma=0.5$)} \\
    \hline
    \multicolumn{3}{c}{Dust}\\
    \hline
    Reddening Law & Calzetti & Fixed \\
    $A_V$ & $(0.0,2.0)$ & Uniform \\
    \hline
    \multicolumn{3}{c}{Nebular}\\
    \hline
    $t_{\mathrm{BC}}$ & 20\,Myr & Fixed \\
    $\log U$ & $(-3.5,-1.0)$ & Uniform \\
    \end{tabular}
    \caption{
    The model components used for the 2D SED fitting with \bagpipes.
    The width of the continuity prior is broader than that used in \cite{leja_how_2019}, to allow for greater variation in the SFH \citep{tacchella_stellar_2022}.
    Any parameters not explicitly listed here were left at their default values.}
    \label{tab:bagpipes_components}
\end{table}

Here we introduce a new method, similar in concept to the one outlined by \citet[][with the code itself termed \textsc{sleuth} in \citealt{estrada-carpenter_metal-poor_2025}]{estrada-carpenter_when_2024}, but implementing a different procedure to construct template spectra.
We use \bagpipes\footnote{\href{https://bagpipes.readthedocs.io/en/latest/}{\texttt{bagpipes.readthedocs.io}}, version 1.3.4} \citep{carnall_inferring_2018} to perform 2D SED fits to the photometry, with the model components summarised in Table~\ref{tab:bagpipes_components}.
We bin the photometry to a target $\sn=10$ in the JWST/NIRCam F150W band via a nearest-neighbours colour scheme \citep{abdurrouf_understanding_2017,estrada-carpenter_when_2024}, which minimises variation of the shape of the SED within each bin.
Starting with the pixel with the highest \sn, this finds the closest matching pixel in colour space, derived as the pair-wise difference between all combinations of the input filters, and adds this to the current bin.
We do not require these pixels to be connected, and hence the bins to be contiguous, unlike the original scheme proposed in \cite{abdurrouf_understanding_2017}.
After reaching the target \sn, a new bin is started with the next highest \sn\ pixel not yet included, and the process is repeated until all pixels have been assigned to a bin.
The median number of bins constructed via this method was 39 for our sample. 
We adopted an additional 5\% uncertainty on the fluxes, to account for systematic variations between the filters \citep[e.g.][]{boquien_cigale_2019,watson_unveiling_2025}.
In the default nested sampling method used by \bagpipes\ \citep{feroz_importance_2019}, the entire parameter space is explored for every SED fit in order to find the optimum solution.
As this is both highly time-consuming,
and inefficient when used for numerous sources occupying comparable regions of the parameter space, we implement an alternative Bayesian grid sampler, similar to the approach taken by other SED fitting codes \citep{boquien_cigale_2019,iyer_nonparametric_2019}.
This has been utilised previously in \cite{benotto_spatially_2026} to derive spatially-resolved properties of some of the most extended galaxies in the same cluster, with comparable accuracy to the original \textsc{MultiNest} sampler when a sufficiently large number of samples are generated (see also the discussion in Appendix A of \citealt{sorba_spatially_2018}).
Here, we construct a model atlas containing $10^6$ samples, as a compromise between speed and accuracy of the SED fit.

After performing the 2D SED fit, we draw a number of samples from the joint posterior distribution in each bin.
These are forward-modelled onto each combination of grism (GR150R, GR150C) and filter (F115W, F150W, F200W), following the morphology and location of the bin in the direct image.
We show in Fig.~\ref{fig:forward_modelling_example} an illustration of the method for one of the galaxies in our sample (ID 3070, $z=1.34$).
A non-negative least squares optimisation is then run to find the optimum combination of all forward-modelled template spectra (Fig.~\ref{fig:forward_modelling_example}D), fitting to all grisms and filters simultaneously.
In the same manner as in the standard \grizli\ reduction, these best-fit models are used to produce continuum-subtracted emission line maps.
The best-fit spectral templates are recalculated without the nebular emission from a particular line (or group of lines, e.g. $\Halpha+\Nii$), then subtracted from the observed grism data in both orientations.
These data are then aligned and drizzled onto a regular North-aligned pixel grid, at a scale of 0\farcs06/pixel.
Consistently with \grizli, the reported integrated line fluxes are therefore the sum of the best-fit model emission fitted in all regions, whereas the line maps are the observed data with the continuum model subtracted.
Properties derived from the initial 2D SED fit, such as stellar mass and dust attenuation, are also reprojected and aligned onto the same grid.

The success of this multiregion method relies on the assumption that the model spectra generated from fits to the photometry are an accurate representation of the observed spectrum.
Whilst an in-depth discussion of SED template mismatch is out of scope of this analysis \citep[we refer the interested reader to analyses such as][]{clausen_performance_2025, luberto_physically-motivated_2025}, we note that this means any unexpected observations outside theoretical predictions will not be fitted properly.
Specific to this analysis, Balmer line flux ratios violating Case B recombination do not lie in the parameter space probed by any current SED fitting code.
We therefore explicitly test the use of a modified set of templates in Appendix~\ref{app:case_b_template_validity}, where \Hbeta\ is artificially boosted by 10\% and 25\% relative to \Halpha, concluding that this has negligible impact on our results.
Similarly, we do not include an AGN component in the \bagpipes\ models at present, although this will be explored in future works.

To minimise the risk of template mismatch in general, we introduce here an extra sampling step.
In addition to the template spectra generated by sampling from the priors of a given region, we also include a number of spectra generated from other (non-contiguous) regions.
In theory, given unlimited computing resources, one could simply fit to the full (discretised) posterior distribution in each bin simultaneously, in order to obtain the best-fit model.
In practice, as RAM is finite and the optimisation routine scales non-linearly with the number of free parameters, it is far more efficient to draw a limited number of posterior samples $\nsamples$, and perform iterations $\niters$ until convergence is achieved.
This assumes that enough posterior samples are drawn in each iteration, that the true spectrum in each region can be constructed from the templates to within the observational uncertainties.
In other analyses \citep{estrada-carpenter_when_2024,estrada-carpenter_metal-poor_2025}, four samples were drawn per region, with several thousand iterations.
For our analysis, we draw five samples from the priors of each region, and an additional sample from the priors of another five regions, for a total $\nsamples=10$ per region.
We iterate only 100 times, with the motivation for this choice discussed further in Appendix~\ref{app:nnls_optimisations}.
This is sufficient that even the largest galaxy in our analysis (ID 1853), with 873 regions, could be fit in under 30 minutes using a laptop CPU, with 32 GB of RAM.

\subsection{PSF matching} \label{method:psf_matching}

It is crucial to note that the NIRISS PSF changes (slowly) over the full wavelength range probed here.
In order to ensure robust comparisons when comparing pixel-level details between emission line maps across different galaxies, at different redshifts, it is necessary to convolve all maps to a common resolution.
We used \textsc{stpsf} \citep{perrin_simulating_2012} to construct a monochromatic PSF at 2.2\textmu m, the approximate long-wavelength limit of the F200W filter, with all model parameters inferred from the date of observation.
This target PSF was rotated based on the \textsc{pa\_aper} FITS header keyword, and reprojected to match the pixel scale of the drizzled line maps.
For each emission line map in our sample, we then repeated this procedure to construct a monochromatic PSF at the observed wavelength of the line.
We used \textsc{pypher} \citep{boucaud_convolution_2016} to derive a kernel matching the observed PSF to the target, and used the \textsc{convolve()} function from \textsc{astropy} to convolve this with the relevant emission line map.

\subsection{Lensing corrections} \label{method:lensing_corrections}

As the GLASS observations targeted the strong lensing cluster Abell 2744 ($z_{\mathrm{cluster}}=0.306$), all of the galaxies in this analysis are magnified to some degree.
We correct all fluxes and derived properties for this using the lensing model of \cite{bergamini_new_2023}.
The magnification $\mu$ of our sample ranges from $\mu=1.5$ (ID 1069) to $\mu=8.9$ (ID 867, also ID 2.1 in \citealt{mahler_strong-lensing_2018}), with a median value of $\mu=2.4$.
We apply a single correction factor per galaxy, and do not correct the emission line maps for spatial distortions arising from the cluster lensing, and present them throughout at a scale of $0\farcs06$/pixel.
Where mentioned, physical scales are corrected for lensing by dividing by a factor $\sqrt{\mu}$, giving a pixel scale for galaxies in this analysis ranging from 170 to 400 parsec/pixel.

\subsection{Line flux corrections} \label{method:line_flux_corrections}

At the spectral resolution of NIRISS ($R\sim150$), \Halpha\ is blended with the nearby \Niiblue\ and \Niired\ lines, which have a theoretical flux ratio of 1:3 respectively for star-forming galaxies \citep{osterbrock_astrophysics_2006}.
Direct measurements of the ``\Halpha'' line are therefore an upper limit on the true \Halpha\ flux.
The exact ratio of \Nii/\Halpha\ is a function of the gas-phase metallicity, for which we have no consistent measurement available (see \citet{estrada-carpenter_metal-poor_2025} at lower redshifts; \citealt{he_early_2024,acharyya_spatially_2026}; K. Nedkova et al. \textit{in prep.} for further discussion of metallicity measurements using NIRISS at Cosmic Noon).
We have instead adopted the
empirical correction from \cite{faisst_empirical_2018}, which essentially parameterises the stellar mass-gas-phase metallicity relation as a function of redshift, calibrated up to $z<2.3$.
We calculate these corrections for each galaxy based on the integrated measurements of the stellar mass from the 2D SED fits, and apply them uniformly to measurements of the \Halpha\ emission complex, including the resolved emission line maps.

We note that the application of these factors as a global correction to all bins within each galaxy will smooth over spatial variations in metallicity, and hence the \Nii/\Halpha\ ratio.
In \cite{estrada-carpenter_metal-poor_2025}, the authors instead correct for the flux contribution from \Nii\ in a spatially-resolved manner, using the empirical relation from \cite{sanchez_integral_2012}, which parameterises \Nii/\Halpha\ as a function of the rest-frame $(B-V)$ colour.
In Appendix~\ref{app:flux_corrections}, we test both the \cite{faisst_empirical_2018} and \cite{sanchez_integral_2012} relations using four galaxies in our sample with archival high-resolution NIRSpec observations \citep{treu_glass-jwst_2022,mascia_glass-jwst_2024}, and find \cite{faisst_empirical_2018} to be significantly more accurate.
We therefore consider the loss of spatial variation in the correction factor to be a reasonable compromise for our analysis, although stress that this represents a significant limitation on both this and other works relying on JWST/NIRISS measurements of \Halpha.

We assume here that all emission is generated by ionisation from star formation.
One potential additional contribution to the Balmer emission comes from diffuse ionised gas \citep[DIG,][]{belfiore_tale_2022}, which can account for a significant fraction of the \Halpha\ emission in local star-forming galaxies.
However, at Cosmic Noon, the fraction of DIG, $f_{\mathrm{DIG}}$, is expected to be minimal \citep{shapley_mosdef_2019}.
Whilst there exist a number of locally-calibrated corrections for $f_{\mathrm{DIG}}$, many of these rely on measurements of the dust-corrected \Halpha\ flux \citep{sanders_biases_2017, tomicic_gasp_2021}, which we eschew due to the correlated uncertainties this introduces into our analysis.
Alternative methods, such as the WHaN diagram, which uses the equivalent width of \Halpha\ to separate DIG-dominated regions \citep{cid_fernandes_alternative_2010, tacchella_h_2022,peluso_interplay_2025}, do not reject any additional regions beyond those already masked out (see Section~\ref{results:case_b:resolved}).
We note that stellar absorption of the Balmer lines is already accounted for in the continuum modelling procedure (Section~\ref{method:multiregion}), and thus does not require a separate correction to be applied.

\section{Results} \label{results}

\subsection{Line flux comparison} \label{results:line_flux_comparison}

\begin{figure}
    \centering
    \includegraphics[width=\columnwidth]{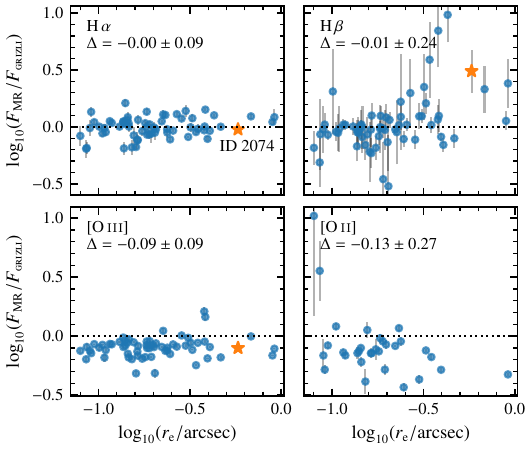}
    \caption{
    The ratio of the integrated emission line fluxes as measured by our multiregion fitting method ($F_{\rm{MR}}$) and \grizli\ ($F_{\grizli}$), against the effective radius of the galaxy, \re.
    The plotted errorbars have been added in quadrature from the two measurements, and the median and standard deviation of the offset are inset for each line.
    We denote the position of ID 2074 (see Fig.~\ref{fig:demonstrate_subtraction}) as an orange star.
    }
    \label{fig:grizli_vs_multiregion_integrated}
\end{figure}

We present in Fig.~\ref{fig:grizli_vs_multiregion_integrated} a comparison of the integrated line fluxes as measured by \grizli\ and our own multiregion fitting method, for four rest-frame optical lines.
For all galaxies, we find substantially lower $\chi^2$ values for the multiregion fits, indicating a better fit to the data.
Although \Halpha\ and \Hbeta\ display no systematic shift, we see considerably higher scatter between the measurements of \Hbeta, with this discrepancy increasing with the effective radii (\re) of the galaxies in our sample, measured from the NIRISS F200W imaging using \textsc{sep} \citep{bertin_sextractor_1996, barbary_sep_2016}.
For \Oii, we find that \grizli\ systematically returns higher estimated line fluxes than our method.
We attribute this to different treatment of the continuum level around the Balmer break, which has also been found when using 1D spectral fitting methods with NIRISS data (K. Nedkova et al., \textit{in prep.}).

For \Oiii, we similarly find higher line fluxes when fitting with \grizli.
Combined with the scatter in \Hbeta, we attribute this to the ``shadowing'' effect of the typically brighter \Oiii\ lines (\citealt{he_early_2024}; \citetalias{watson_glass-jwst_2025}), whereby the spectral and spatial overlap prevents a robust extraction of either \Oiii\ or \Hbeta.
This self-contamination appears to lead to the stronger line (\Oiii) being assigned a portion of the flux from the weaker line (\Hbeta), when only a single spectrum is fit to extended sources.
Similar effects have been observed before, regarding contamination of \Sii\ emission by \Halpha\ \citep{acharyya_spatially_2026}, demonstrating the vital importance of multi-region modelling.

\begin{figure}
    \centering
    \includegraphics[width=\columnwidth]{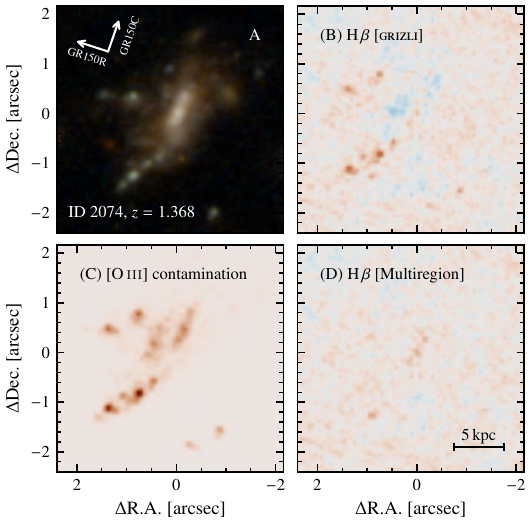}
    \caption{
    (A) RGB image of the galaxy ID 2074.
    (B) The \grizli-derived \Hbeta\ emission line map, with the \grizli-modelled contamination from nearby emission lines subtracted.
    (C) The multiregion-derived models for \Oiiiblue\ and \Oiiired, at the drizzled wavelength of \Hbeta. Notice that most prominent peaks in (B) correspond precisely to the location of the \Oiii\ emission, which has been offset along the two trace directions (inset).
    (D) The multiregion-derived emission line map for \Hbeta, where contamination from \Oiii\ has been correctly subtracted.
    These maps are shown before the binning scheme is applied.
    }
    \label{fig:demonstrate_subtraction}
\end{figure}

We illustrate this effect in Fig.~\ref{fig:demonstrate_subtraction} for one of the larger galaxies in our sample (ID 2074).
This source contains multiple off-centre star-forming clumps, visible in the constructed RGB image in Fig.~\ref{fig:demonstrate_subtraction}A, and a central component dominated by continuum emission.
The net effect of this is that when using the direct image as a reference for the morphology of all emission lines, the line flux is both underestimated in the outskirts, and overestimated in the centre.
This introduces considerable contamination when extracting relatively fainter lines, such as \Hbeta, shown in Fig.~\ref{fig:demonstrate_subtraction}B.
We illustrate in Fig.~\ref{fig:demonstrate_subtraction}C that this extracted line map is dominated by the residual emission from \Oiii, plotting the multi-region model for \Oiii, shifted along the grism trace directions at the wavelength of \Hbeta.
The most prominent peaks in the \grizli-derived \Hbeta\ map correspond precisely to the expected locations of the \Oiii\ emission.
We show instead in Fig.~\ref{fig:demonstrate_subtraction}D the multiregion-derived \Hbeta\ line map, where the substantially fainter peaks are now correctly aligned with the clumps visible in the direct imaging.

\subsection{Case B violation} \label{results:case_b}

We investigate here the extent to which Case B recombination can be considered applicable to our measurements of the Balmer decrement, for both the integrated and spatially-resolved samples.
Throughout this analysis, we assume the canonical value of $(\Halpha/\Hbeta)_{\mathrm{intr}}=2.86$ from \cite{osterbrock_astrophysics_2006}, corresponding to a nebular temperature $T=10^4$\,K and an electron density $\edens = 10^2\,\mathrm{cm}^{-3}$.
Following \cite{dominguez_dust_2013}, we also convert the observed values of the Balmer decrement into a nebular colour excess \EBVneb, using a reddening curve \klam{}:
\begin{equation}
    \EBVneb = \frac{2.5}{\klam{\Hbeta} - \klam{\Halpha}} \log_{10}\left( \frac{(\Halpha/\Hbeta)_{\mathrm{obs}}}{(\Halpha/\Hbeta)_{\mathrm{intr}}} \right),
\end{equation}
where we adopt the \cite{calzetti_dust_2000} extinction curve to obtain
\begin{equation}
    \frac{2.5}{\klam{\Hbeta} - \klam{\Halpha}}  = 1.97\,.
\end{equation}

\subsubsection{Integrated balmer decrement} \label{results:case_b:integrated}

\begin{figure}
    \centering
    \includegraphics[width=\columnwidth]{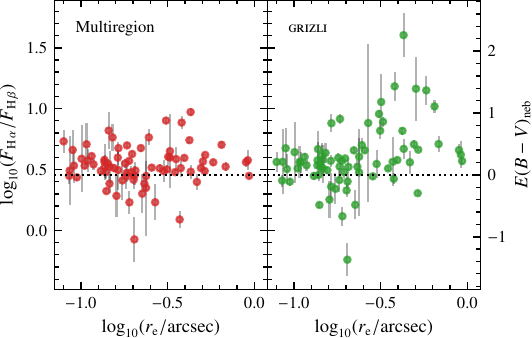}
    \caption{
    The Balmer decrement \Halpha/\Hbeta\ from integrated measurements using both our multiregion method and \grizli, on a logarithmic scale, against the logarithm of the galaxy effective radius. 
    On the right-hand side of the axes, we show the equivalent nebular colour excess, \EBVneb.
    The dotted line indicates the Case B limit, or equivalently $\EBVneb=0$.
    }
    \label{fig:grizli_vs_multiregion_integrated_bd}
\end{figure}

In Fig.~\ref{fig:grizli_vs_multiregion_integrated_bd}, we show the distribution of the integrated Balmer decrement derived from both fitting methods, against the effective radius measured on the stacked detection image.
Uncertainties on the Balmer decrement are calculated by summing the uncertainties on the fluxes in quadrature.
As expected from Section~\ref{results:line_flux_comparison}, significant discrepancies are visible in Fig.~\ref{fig:grizli_vs_multiregion_integrated_bd} between the more extended galaxies, where the underestimate by \grizli\ of the \Hbeta\ emission results in an overestimate of the nebular attenuation.
We calculate the mean Balmer decrement from \grizli\ as $\Halpha/\Hbeta=4.0$, consistent with the independently-derived value of 4.06 from \cite{pang_glass-jwst_2026}, using an almost identical sample, but considerably higher than the mean of $\Halpha/\Hbeta=3.4$ derived from the multiregion fits.
For the rest of this analysis, we therefore use only the measurements from the multiregion fit, unless otherwise specified.
From these measurements, we find 5 galaxies with a derived $\EBVneb<0$ at $\geq1\sigma$, corresponding to $\sim7$\% of the sample.
This does not vary even if we explicitly include templates with $\Halpha/\Hbeta$ ratios outside the regime described by Case B recombination (see Appendix~\ref{app:case_b_template_validity}), indicating that this fraction has not been artificially suppressed by our fitting method and template selection.

\begin{figure}
    \centering
    \includegraphics[width=\columnwidth]{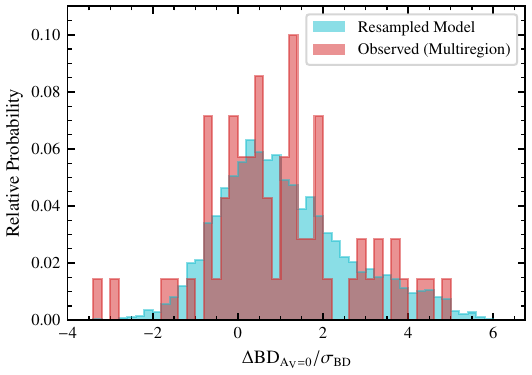}
    \caption{
    The distance of the integrated measurements of the Balmer decrement (BD) from the Case B limit ($\Delta\mathrm{BD}_{A_V=0}$), normalised by the uncertainty $\sigma_{\mathrm{BD}}$.
    We show both the observed values, and the median model after resampling 1000 times, assuming no measurements are intrinsically below the Case B limit.
    }
    \label{fig:BD_distance_model_dustfree_integrated}
\end{figure}

We test the statistical significance of this proportion, against the null hypothesis that the intrinsic Balmer decrement agrees with Case B recombination theory, and deviations from this are purely due to observational uncertainties.
We perform a bootstrap procedure, resampling from the observed Balmer Decrements $F_{\Halpha}/F_{\Hbeta}$ (hereinafter abbreviated as \BD) assuming a Gaussian distribution with standard deviation equal to the 1$\sigma$ uncertainties.
For all measurements in which $\Halpha/\Hbeta<2.86$, we set the value of \Hbeta\ to $\Halpha/2.86$, but do not rescale the measurement uncertainties.
From this, we resample 1000 times, and show the resulting distribution against the observed in Fig.~\ref{fig:BD_distance_model_dustfree_integrated}. We find the median estimate of the Case B violation fraction to be $5.3\pm2.1\%$, compared against the observed $6.6\%$.
We therefore conclude that this proportion of galaxies below the Case B limit appears to be driven solely by our observational uncertainties, rather than any physical explanation.

\subsubsection{Resolved emission} \label{results:case_b:resolved}

While useful in its own right, integrated analyses cannot fully reveal the variation in dust attenuation within galaxies \citep[e.g.][]{nelson_spatially_2016,greener_sdss-iv_2020, lee_spatially_2025, barisic_msa-3d_2025}.
We therefore also investigate the spatially-resolved properties of our sample.
As noted in other previous analyses \citep[e.g.][]{he_early_2024, acharyya_spatially_2026}, the \sn\ of the GLASS/NIRISS data is often insufficient for comparisons at the scale of individual pixels, and it is thus necessary to bin to achieve the required \sn.

Our results show a small dependence on the binning scheme used.
We test the weighted Voronoi tessellation algorithm of \cite{diehl_adaptive_2006}, as implemented in \textsc{vorbin} \citep{cappellari_adaptive_2003}, the newer \textsc{powerbin} algorithm proposed in \cite{cappellari_powerbin_2025}, and the nearest-neighbour colour binning scheme discussed earlier in Section~\ref{method:multiregion}.
For all methods, we target a $\sn=6$ in \Halpha.
Assuming that the uncertainty on the resolved emission line fluxes is dominated by background noise, and is therefore comparable between \Halpha\ and \Hbeta\ for any given galaxy (see Appendix~\ref{app:background_levels}), this would give an approximate \Hbeta\ detection at a $\sn\gtrsim2$ in the Case B recombination limit.
We place some additional constraints on the resulting bins, to remove spurious detections and low-quality data.
We require \Hbeta\ to be detected at a $\sn\geq1$, and that both \Halpha\ and \Hbeta\ exceed the background RMS, measured in Appendix~\ref{app:background_levels}.
This latter requirement is to mitigate against spurious detections in both single-spaxel bins, and large bins near the low \sn\ edges of the segmentation map.

\begin{figure}
    \centering
    \includegraphics[width=\columnwidth]{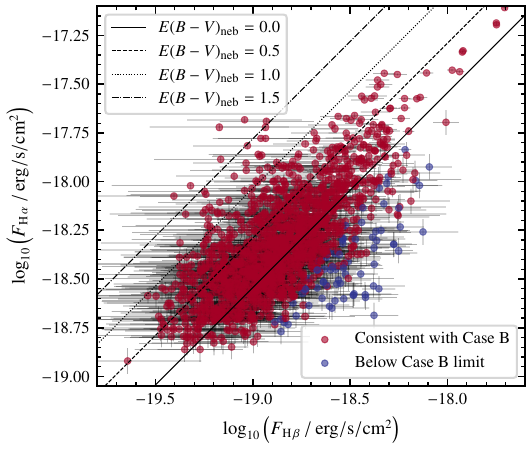}
    \caption{
    The distribution of \Halpha\ fluxes against \Hbeta\ fluxes, as measured in 1450 individual bins across 76 sources in our analysis.
    We overlay lines indicating the expected ratios of \Halpha/\Hbeta\ for a range of values of nebular attenuation, assuming Case B recombination.
    Bins inconsistent with Case B recombination, considering the $1\sigma$ uncertainties on \Halpha\ and \Hbeta, are coloured in blue.
    }
    \label{fig:colour_binning_Ha_Hb}
\end{figure}

Comparing binning schemes, both \textsc{vorbin} and \textsc{powerbin} are constrained to only construct contiguous bins.
Given the relatively low \sn\ of the data here, when combined with the other constraints on shape and size, this often results in bins falling below our minimum \sn\ requirements.
This is reflected in the total number of bins and sources passing the quality checks for each scheme --- binning based on colour yields 1450 bins from 76 sources, \textsc{powerbin} gives 1704 bins from 62 sources, and \textsc{vorbin} creates 1376 bins from only 50 sources.
We therefore adopt the colour binning scheme for the remainder of our analysis, although we wish to stress that our results do not qualitatively change based on this decision (see Appendix~\ref{app:binning_schemes}). 

A significant fraction of our sample (27 galaxies) contains no detection of \Hbeta\ above the background level, thus placing a lower limit on the derived Balmer decrement and nebular reddening.
Whilst these data may be useful for some studies, they do not enable us to assess the likelihood of Case B recombination violation, and place negligible constraints on any other scaling relations in this analysis.
Henceforth, we remove these galaxies from the sample under consideration in all further parts of this analysis.

We show the resulting binned \Halpha\ and \Hbeta\ fluxes in Fig.~\ref{fig:colour_binning_Ha_Hb} using the colour binning scheme.
Across the full ensemble of bins, we find a median $\sn=9.6$ in \Halpha\ and $\sn=2.4$ for \Hbeta.
Of these, 76 individual bins ($5.3\%$ of the total) have \Halpha/\Hbeta\ ratios lower than the dust-free Case B prediction, accounting for the $1\sigma$ observational uncertainties.
This is a similar level to that observed using the integrated measurements in Section~\ref{results:line_flux_comparison}.

Considering the relatively low \sn\ of the resolved measurements, we test the statistical significance of this fraction in the same manner as in Section~\ref{results:case_b:integrated}.
With 1000 iterations, we find an estimated Case B violation fraction of $5.2\pm0.5\%$.
Therefore, across the entirety of our resolved sample, the fraction of regions we find to be below the Case B recombination limit is wholly consistent with a population at the limit, combined with our measurement uncertainties.

\subsection{Redshift dependence} \label{results:redshift_dependence}

\begin{figure}
    \centering
    \includegraphics[width=\columnwidth]{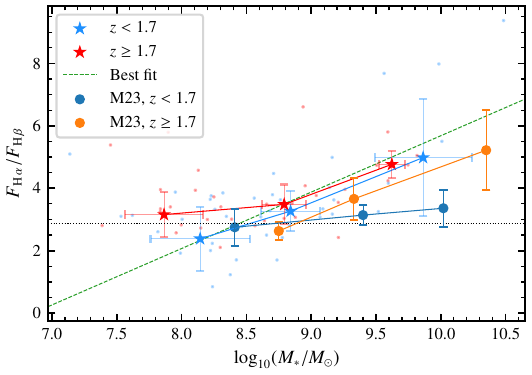}
    \caption{
    The integrated Balmer decrement as a function of total stellar mass, split into two redshift bins.
    We show the median values and their uncertainties in each of three stellar mass bins, plotted as stars.
    Circles denote the results from \cite{matharu_first_2023}, and the green dashed line shows the best-fit linear relation to our full sample.
    }
    \label{fig:balmer_decrement_vs_mass}
\end{figure}

In Fig.~\ref{fig:balmer_decrement_vs_mass}, we show the integrated measurements of the Balmer decrement as a function of total stellar mass.
We divide our sample into two redshift bins, $z<1.7$ (42 galaxies) and $z\geq1.7$ (32 galaxies).
For each redshift subsample, we calculate the median in three stellar mass bins, $\logstellmass<8.5$, $8.5\leq\logstellmass<9.5$, and $\logstellmass\geq9.5$.
As a comparison, we also plot the results from \cite{matharu_first_2023}, which used JWST/NIRISS data of comparable depth and wavelength coverage, and split the sample in redshift in the same manner.
Although our results are generally consistent within the uncertainties on the median, we do not observe the same cross-over between the redshift bins at low masses.
At fixed stellar mass, we also find higher values of the Balmer decrement than \cite{matharu_first_2023}, regardless of redshift.
The most probable explanation for this, aside from sample selection bias, is the difference in the wavelength and trace calibration used during the data reduction stage.
As we demonstrate in Appendix~\ref{app:pipeline_differences}, across this redshift range the \cite{matharu_updated_2022} calibrations returned systematically higher values of \Hbeta, which would in turn lead to lower measurements of the Balmer decrement.

In the majority of previous analyses, it has been shown that the dust content of a galaxy correlates primarily with its stellar mass, with little or no redshift dependence until $z\gtrsim5$ \citep{dominguez_dust_2013,price_direct_2014,shapley_mosfire_2022,matharu_first_2023,woodrum_jades_2025,karthikeyan_balmer_2026,alavi_uv_2026}.
On the other hand, previous works making use of JWST/NIRISS data have reported contrasting results, either finding no correlation with stellar mass \citep{pirzkal_next_2024}, or that the correlation varies as a function of redshift \citep{matharu_first_2023, ren_early_2026}.
Here, our results agree with the majority of the literature, finding both a clear correlation between the integrated Balmer decrement and stellar mass of the galaxy, and also
finding no measurable evolution in the relation between our two redshift bins.
As such, we show in Fig.~\ref{fig:balmer_decrement_vs_mass} the best-fit linear relation to all of our measurements, independent of redshift, parameterised as:
\begin{equation}
    F_{\Halpha}/F_{\Hbeta} = (1.81\pm0.23)\times(\log M_*-9.0) + (3.88\pm0.13).
\end{equation}
We also note that at lower masses ($\logstellmass\lesssim8.5$), this relationship appears to plateau, consistent with previous analyses \citep{dominguez_dust_2013,shapley_mosfire_2022,alavi_uv_2026}, indicating little or no attenuation at these stellar masses.

\subsection{Scaling relations} \label{results:scaling_relations}

\begin{figure}
    \centering
    \includegraphics[width=\columnwidth]{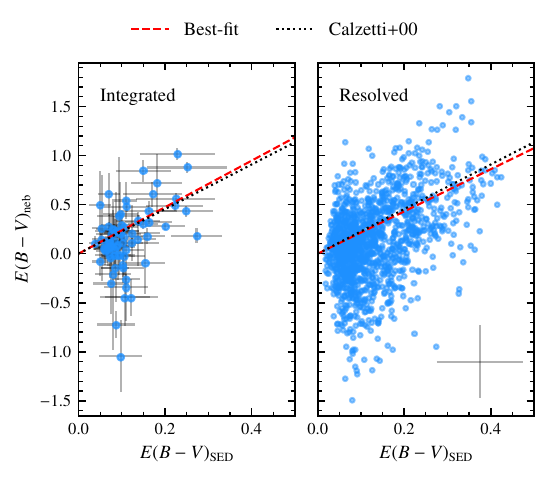}
    \caption{
    (A) The correlation between the nebular reddening \EBVneb\ measured from the Balmer decrement \Halpha/\Hbeta, and the the stellar continuum reddening \EBVsed. 
    We show in (A) the integrated measurements, and in (B) the resolved measurements for the full sample.
    The best-fit linear relation is plotted as a red dashed line, with the local relation derived by \cite{calzetti_dust_2000} shown as a black dotted line.
    The median uncertainties are shown inset for the resolved sample.
    }
    \label{fig:ebv_neb_vs_sed}
\end{figure}

As discussed in Section~\ref{sec:intro}, the observed dust attenuation of nebular regions is typically higher than that of stellar regions.
In large samples of local galaxies \citep[e.g.][]{calzetti_dust_2000,zahid_census_2012,lin_variant_2020} this correlation is typically formulated in terms of the reddening as $\EBVsed=0.44\times\EBVneb$, where \EBVsed\ indicates the stellar continuum reddening, as measured from the SED using \textsc{bagpipes} (see Section~\ref{method:multiregion}).
We show the equivalent relation for our sample in Fig.~\ref{fig:ebv_neb_vs_sed}.
For both the integrated and resolved measurements, we do not observe any significant deviation from the local relation derived by \cite{calzetti_dust_2000}.
We derive the best-fit integrated relation using \textsc{lmfit}, weighted by the uncertainties, as $\EBVsed=(0.42\pm0.05)\times\EBVneb$, and $\EBVsed=(0.46\pm0.02)\times\EBVneb$ for the resolved.
Thus, despite the significantly higher scatter on the resolved sample, we do not find any evidence that this follows a different scaling relation to the integrated properties.

There is no clear consensus within the literature as to how this factor relating \EBVsed\ to \EBVneb\ varies as a function of redshift.
\cite{price_direct_2014} derive a factor of 0.53 at $z\sim1.4$, \cite{shivaei_investigating_2015} adopt 
$\EBVsed=\EBVneb$ at $z\sim2$, \cite{tsujita_alpine-cristal-jwst_2026} derive a factor $0.51$ at $z\sim5$, whilst \cite{woodrum_jades_2025} find a median factor 0.32 at $3<z<7$.
Our results favour a minimal evolution with redshift from the local universe at least up to $z\sim2$, although we note that the relation shown in Section~\ref{results:scaling_relations} has considerable scatter, and the exact method of determining the stellar continuum reddening may have a significant impact on the derived relation.
In particular, the normalisation of the adopted reddening curve \citep[e.g.][]{cardelli_relationship_1989,calzetti_dust_2000} can lead to systematic shifts, a full discussion of which is beyond the scope of this analysis (see for example \citealt{battisti_average_2022}).

\subsection{Differential reddening} \label{results:differential_reddening}

At higher redshift, many observations lack measurements of multiple Balmer lines, necessary to derive direct measurements of the nebular attenuation.
However, accurate measurements of the stellar attenuation can be far more easily derived from SED fits to the broad-band photometry.
Whilst Section~\ref{results:scaling_relations} and many other analyses demonstrate that the two quantities are correlated, the scatter is considerable, and limits its application. 
In order to obtain a robust empirical relationship between the two, we examine how the differential nebular-stellar reddening, $\Delta\EBV=\EBVneb-\EBVsed$, correlates with other quantities that can be derived from the SED.

\subsubsection{Integrated relations} \label{results:differential_reddening:integrated}

\begin{figure}
    \centering
    \includegraphics[width=\columnwidth]{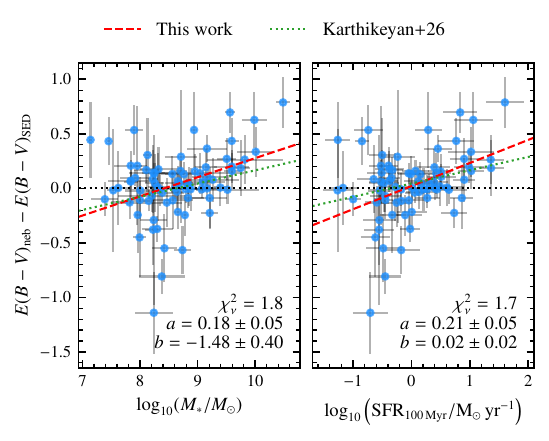}
    \caption{
    (A) The correlation between integrated measurements of the differential nebular-stellar attenuation, and the total stellar mass of the galaxy.
    The best-fit linear relation is indicated by the red dashed line, with the parameters and uncertainties inset.
    (B) As in (A), but plotted against the average SFR over the last 100\,Myr, measured from SED fitting.
    In each plot we overlay the corresponding best-fit relation from \cite{karthikeyan_balmer_2026} as a dotted line for comparison.
    }
    \label{fig:stellar_vs_neb_ebv_integrated}
\end{figure}

For the integrated measurements, we consider the differential reddening as a function of both the stellar mass and the star-formation rate averaged over the last 100\,Myr, adopting the shorthand $\log\mathrm{SFR}_{100}\equiv\log_{10}\left(\mathrm{SFR_{100\,Myr}}/\sfrunits\right)$.
The latter is readily available from a range of SED fitting codes, and is robust as long as the photometry includes the rest-frame UV regime (as mentioned in Section~\ref{method:ancillary_data}, the rest-frame coverage extends to $\lesssim0.2$\,\textmu m for this analysis).
We show the best-fit relations in Fig.~\ref{fig:stellar_vs_neb_ebv_integrated}, finding weak positive correlations with $\Delta\EBV$ of the form
\begin{equation}
    \Delta\EBV = (0.18\pm0.05)\logstellmass - (1.48\pm0.40) 
\end{equation}
and 
\begin{equation}
    \Delta\EBV = (0.21\pm0.05)\log\mathrm{SFR_{100}} + (0.02\pm0.02).
\end{equation}
We calculate the Spearman correlation coefficient, finding $\rho=0.27$ for the relation with stellar mass ($p=0.016$), and $\rho=0.33$ for the relation with SFR ($p=0.004$).

As a comparison, we look at the relations derived by \cite{karthikeyan_balmer_2026}, although their redshift range does not overlap.
In their lowest redshift bin, $2.7\leq z<4.0$, they find the differential nebular-stellar reddening to be correlated with mass and SFR as $\Delta\EBV\sim(0.121\pm0.011)\logstellmass$ and $\Delta\EBV\sim(0.122\pm0.011)\log\mathrm{SFR}_{100}$, shown also in Fig.~\ref{fig:stellar_vs_neb_ebv_integrated}.
Neither of these relations are quite consistent with those we derive, which have higher positive gradients, albeit with wider uncertainties.
Qualitatively though, our results agree that the nebular emission has systematically higher attenuation relative to the stellar emission at both higher masses and star formation rates.
Considering two component dust models, such as \cite{charlot_simple_2000}, this would imply that in more massive galaxies, dust is relatively more concentrated within stellar birth clouds than in the diffuse ISM.

\subsubsection{Resolved relations} \label{results:differential_reddening:resolved}

\begin{figure}
    \centering
    \includegraphics[width=\columnwidth]{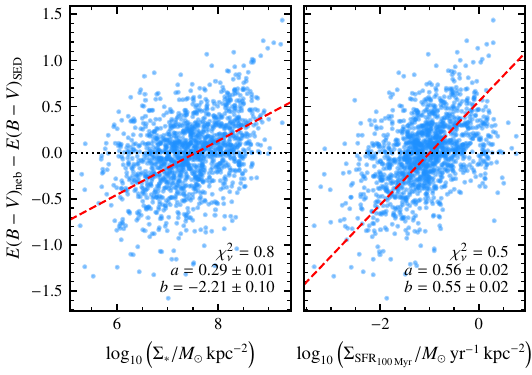}
    \caption{
    (A) The correlation between resolved measurements of the differential nebular-stellar attenuation, and the stellar surface mass density of the corresponding region.
    The red dashed line indicates the best-fitting linear relation, with the parameters and their uncertainties inset.
    (B) As in (A), but as a function of the average SFR surface density over the last 100\,Myr, measured from SED fitting.
    The median errors are shown inset.
    }
    \label{fig:differential_ebv_resolved}
\end{figure}

For the resolved measurements, we investigate correlations with the stellar mass surface density, $\logmstarsurfdens\equiv\log_{10}\left(\Sigma_{M_*}/\Msol\,\mathrm{kpc}^{-2}\right)$, and the SFR surface density $\logsfrsurfdens\equiv\log_{10}\left(\mathrm{\Sigma_{SFR_{100\,Myr}}}/\sfrunits\,\mathrm{kpc}^{-2}\right)$.
We derive the best-fit relations in the same manner as in Section~\ref{results:differential_reddening:integrated}, with the results shown in Fig.~\ref{fig:differential_ebv_resolved}.
We find 
\begin{equation}
    \Delta\EBV_{\mathrm{res}} = (0.29\pm0.01)\logmstarsurfdens - (2.2\pm0.1)
\end{equation}
and 
\begin{equation}
    \Delta\EBV_{\mathrm{res}} = (0.56\pm0.01)\logsfrsurfdens + (0.55\pm0.01).
\end{equation}
The Spearman rank correlation coefficients for these quantities are $\rho=0.38$ and $\rho=0.45$ respectively, with $p<<0.001$ for both.
Through a comparison of the normalised Bayesian Information Criterion (see Appendix~\ref{app:case_b_template_validity}), we find that the relation with \sfrsurfdens\ is preferred over the relation with \logmstarsurfdens, although the evidence in favour is relatively weak ($\Delta\mathrm{BIC}\sim0.5$). 
We also test correlations with the specific star formation rate, $\mathrm{sSFR=SFR}/M_*$, but find this provides poorer constraints than either parameter individually, due to the compounding uncertainties on the SED-derived properties.

Similar results have been obtained before for local galaxies.
In \cite{lee_spatially_2025}, the authors find a strong correlation between the differential attenuation (parameterised in a different manner) and both \sfrsurfdens\ and sSFR.
However, their measurements of SFR were derived from the dust-corrected \Halpha\ flux, and so traces different timescales of star formation to that used in Fig.~\ref{fig:differential_ebv_resolved}.
Likewise, in deriving a parameterisation of the resolved nebular attenuation $A_V$, \cite{mailvaganam_predicting_2026} found the strongest predictor to be \sfrsurfdens, although they did not consider the stellar attenuation as a possible parameter.
At higher redshift, using $\sim0.6$kpc/pixel resolved maps, \cite{tsujita_alpine-cristal-jwst_2026} examined the fractional reddening, $\EBVneb/\EBVsed$. 
In contrast to both the local analyses and our own results, they found the strongest correlation with the sSFR over 10\,Myr, with no significant correlation with $\sfrsurfdens$.
Therefore, to assess the extent to which our relation with \sfrsurfdens\ reflects resolved physical trends, rather than simply sample variation, we refit this individually for the 19 galaxies which each contain at least 20 bins.
We find that these individual relations are overall consistent to that shown in Fig.~\ref{fig:differential_ebv_resolved}, with a median gradient of $\Delta\EBV_{\mathrm{res}}\propto(0.63\pm0.30)\logsfrsurfdens$.

\section{Discussion} \label{discuss}

\subsection{Case B recombination} \label{discuss:case_b}

A number of studies have observed galaxies with Balmer decrements below the theoretical Case B recombination limit, including some analyses in which these sources account for up to 40\% of the total sample (e.g. \cite{pirzkal_next_2024} with JWST/NIRISS, \cite{shapley_jwstnirspec_2023, sandles_jades_2024, sun_mysteriously_2025, woodrum_jades_2025,llerena_extreme_2026} with JWST/NIRSpec, and \cite{alavi_uv_2026} with Keck/MOSFIRE).
However, there is considerable variation in how this fraction is derived.
\cite{sandles_jades_2024} consider the measurement uncertainties on the Balmer decrement, finding that this accounts for all but two of the seemingly incongruent galaxies, and similarly 89\% of the sample in \cite{llerena_extreme_2026}.
On the other hand, \cite{sun_mysteriously_2025}, reporting the largest proportion of Case B violating galaxies at 40\%, consider only the central measurement of the fluxes when assessing the proportion of Balmer decrements below 2.86.
In contrast, we find only $\sim$7\% of integrated measurements of the Balmer decrement in our sample lie more than 1$\sigma$ below the Case B limit.
We also test this using spatially resolved measurements, finding $\sim$6\% of bins inconsistent with the limit, indicating that this cannot solely be attributed to artefacts with the data processing of extended galaxies.

To some extent, small deviations from the intrinsic Case B limit are expected.
Although the Balmer line ratios are broadly insensitive to electron temperature and density, within the range $5\times10^4<T<3\times10^5$\,K and $10<\edens <500\,\mathrm{cm}^{-3}$, the Balmer decrement varies by $\sim6\%$ \citep{sandles_jades_2024}.
In some analyses, with direct measurements of the electron temperature or density, it is possible to determine the exact value of the intrinsic Balmer decrement, and consider the physical mechanisms involved \citep[e.g.][]{scarlata_universal_2024}.
Here we adopt a more straightforward observational argument.
Given the \sn\ of our observations, it is probable that some galaxies (or invididual bins) will be observed to have $\Halpha/\Hbeta<2.86$, particularly if the intrinsic attenuation in the galaxy is near the dust-free limit.
As shown in Section~\ref{results:case_b}, we have demonstrated that this hypothesis holds for the GLASS data.
The apparently incongruent galaxies are in fact consistent with the expected scatter from a dust-free population, i.e. galaxies at the Case B limit.
Crucially,
a full consideration of the uncertainties is essential to avoid misleading results.
In our integrated (resolved) sample, 25\% (34\%) of galaxies have Balmer decrements below 2.86 when only looking at the measured values, and this may explain the high proportion reported by \cite{sun_mysteriously_2025}.

We also stress that scatter from measurement uncertainties may not be sufficient to explain all similar observations, although it has already been suggested in some analyses \citep{shapley_jwstnirspec_2023,sandles_jades_2024}.
As first discussed by \cite{scarlata_universal_2024}, in local galaxies there does appear to be a link between the ionisation state and the Balmer decrement, where galaxies with high $O_{32}\equiv\Oiiired/\Oii$ ratios have a higher proportion of sub-Case B Balmer decrements.
This is corroborated by \cite{alavi_uv_2026}, who find 15\% of their sample with $O_{32}<5$ have $\Halpha/\Hbeta<2.86$, compared to $\sim40\%$ at $O_{32}>5$.
We test this with the 32 galaxies in our integrated sample for which \Oii\ is available.
Taking our measurements at face value, we find similar proportions here, with $\sim$10\% of galaxies at $O_{32}<5$ having $\Halpha/\Hbeta<2.86$, increasing to $\sim35\%$ at $O_{32}>5$,
although we caution that as discussed above, these fractions are reduced to negligible amounts when accounting for the uncertainties on our line ratios.

There are still some caveats to this part of our analysis, many of which stem from the NIRISS dataset used.
As discussed in Section~\ref{method:line_flux_corrections}, due to the low resolution of the GR150R/C grisms ($R\sim150$), \Halpha\ and \Nii\ are blended into a single line complex.
We have adopted a single uniform correction factor per galaxy here, smoothing over any internal variation.
Although this may increase the scatter on the relations, we do not consider this to have had a qualitative impact on our analysis.
The median correction applied is $\Nii/(\Nii+\Halpha)=4.4\%$, whilst the largest individual correction applied to any galaxy in our sample is 36\%.
However, we do caution that if \Nii/\Halpha\ and \EBV\ are correlated \citep{battisti_average_2022,ji_need_2023,lee_spatially_2025}, then by applying a global correction we would be overestimating the \Halpha\ flux in the most attenuated bins, and so the true $\Delta\EBV\sim\sfrsurfdens$ relation would likely be flatter.
To obtain a more accurate calibration, is is therefore essential to obtain larger, high \sn\ samples in which we can directly measure the metallicity and thus correct for the spatial variation in \Nii/\Halpha.
Additionally, further research into the optimum correction for the \Nii/\Halpha\ ratio would greatly reduce the uncertainties associated with similar grism-based analyses.

\subsection{Sample completeness} \label{discuss:sample_completeness}

By construction, our sample is unlikely to be fully representative of all galaxies at Cosmic Noon.
The initial redshift selection requires sufficiently large equivalent widths in multiple lines that a secure redshift can be determined.
For the sample here, where \Halpha\ and \Oiii\ are the dominant emission lines, this corresponds to a lower limit in SFR of $\sim0.4$\Msol/yr$^{-1}$ \citep{pang_glass-jwst_2026}.
We have also required a detection of \Hbeta\ when deriving the empirical calibrations, which will bias our sample against highly attenuated galaxies.

With this in mind, for mass or magnitude-limited samples at $z\gtrsim2$, it is unclear how representative the calibrated attenuation corrections in Section~\ref{results:differential_reddening} will be.
However, one of the primary sources of spatially-resolved data at these redshifts will continue to be WFSS observations, such as ongoing surveys with JWST/NIRISS and the upcoming High-Latitude Wide Area Survey with the Roman space telescope \citep{observations_time_allocation_committee_roman_2025}.
For observations such as these, we expect similar caveats to apply to the sample selection and completeness, and as such, our calibrations will be broadly applicable.

\section{Conclusions} \label{conclusions}

This analysis presents an 
updated reduction of the JWST/NIRISS data from the GLASS-JWST ERS programme, 
publicly available at \href{https://doi.org/10.5281/zenodo.20613558}{Zenodo}.
We use an advanced multiregion fitting method to extract
emission line maps from these data, accounting for the spatial variation in emission line ratios and stellar populations, with these codes also publicly available on 
\href{https://github.com/PJ-Watson/niriss-tools}{GitHub}.
We use a sample of 99 galaxies where both \Halpha\ and \Hbeta\ are covered by the grism filters, with \Halpha\ detected at an integrated $\sn>5$, and where the ionising source lies in the star-forming region of the $OHNO$ parameter space \citep{acharyya_spatially_2026}.
Of these, 76 sources have \Hbeta\ detectable above the background level in the drizzled emission line maps.
Our results can summarised as follows:
\begin{itemize}
    \item We demonstrate that the assumption of spatially homogeneous emission in codes such as \grizli\ breaks down for more extended galaxies, leading to systematic offsets in the derived emission line fluxes.
    Due to the proximity of \Oiii, \Hbeta\ is typically underestimated in the largest galaxies, artificially increasing the measured Balmer decrement (\Halpha/\Hbeta).
    \item We find $\sim$7\% of our sample (from both integrated and spatially-resolved measurements) have Balmer decrements more than 1$\sigma$ below the canonical value of 2.86 from Case B recombination.
    This proportion is entirely consistent with the observational uncertainties if all these galaxies had an intrinsic Balmer decrement of $\Halpha/\Hbeta=2.86$, in line with a dust-free population.
    \item We reproduce the known correlation between the Balmer decrement and the total stellar mass of the galaxy, finding no redshift evolution from $z<1.7$ to $z\geq1.7$.
    We fit the entire sample to a linear relation as
    \begin{equation}
        F_{\Halpha}/F_{\Hbeta} = 1.81\times(\log M_*-9.0) + 3.88,
    \end{equation}
    noting that at $\logstellmass\lesssim8.5$, this would indicate little or no dust attenuation at these masses.
    \item Our integrated (resolved) sample displays no deviation from the local scaling relation between \EBVneb\ and \EBVsed\ for starburst galaxies,
    with $\EBVsed=0.44(0.46)\times\EBVneb$ respectively.
    \item The differential nebular-stellar reddening, $\Delta\EBV=\EBVneb-\EBVsed$ displays a statistically significant correlation with both stellar mass and the SFR averaged over the last 100\,Myr.
    \item For resolved measurements of $\Delta\EBV$, we find weak evidence ($\Delta\mathrm{BIC}\sim0.5$) in favour of a correlation with the star formation rate surface density over the stellar mass surface density, which we parameterise as
    $$\Delta\EBV=0.56\sfrsurfdens+0.55.$$
\end{itemize}
These measurements further establish the use of WFSS as an integral tool for spatially-resolved science at Cosmic Noon, with these empirical attenuation calibrations enabling further science at higher redshifts.

\section*{Data availability} \label{data_availability}

The raw data used are available at the MAST, with JWST programme ID 1324.
The redshift catalogue is also available at the MAST, as part of the GLASS-JWST High-Level Science Product collection\footnote{\url{https://archive.stsci.edu/hlsp/glass-jwst}}.
We have made the \grizli-reduced data products available on Zenodo\footnote{\url{https://doi.org/10.5281/zenodo.20613558}}, alongside Python scripts to extract individual galaxies or regions of interest.
The source code for the multiregion fitting is available on GitHub\footnote{\url{https://github.com/PJ-Watson/niriss-tools}}, alongside both example scripts and those used to derive the data in this manuscript. 

\begin{acknowledgements}
    We thank the anonymous referee whose useful comments helped us to improve the manuscript. 
    
    The data were obtained from the MAST at the Space Telescope Science Institute, which is operated by the Association of Universities for Research in Astronomy, Inc., under NASA contract NAS 5-03127 for JWST. 
    Some of the data products presented herein were retrieved from the Dawn JWST Archive (DJA). DJA is an initiative of the Cosmic Dawn Center (DAWN), which is funded by the Danish National Research Foundation under grant DNRF140.
    These observations are associated with programme JWST-ERS-1324. 
    We acknowledge financial support from NASA through grants JWST-ERS-1324.
    We also acknowledge support from the INAF Large Grant 2022 `Extragalactic Surveys with JWST' (PI Pentericci). 
    P. W., B.V. and A.A. are supported by the European Union – NextGenerationEU RFF M4C2 1.1 PRIN 2022 project 2022ZSL4BL INSIGHT. 
    P.W. and B.V. acknowledge support from the INAF Mini Grant `1.05.24.07.01 RSN1: Spatially Resolved Near-IR Emission of Intermediate-Redshift Jellyfish Galaxies' (PI Watson).
    We thank both the GLASS-JWST and PASSAGE teams for many useful discussions on working with NIRISS data.
\end{acknowledgements}

%
\bibliographystyle{aa} 
\bibliography{references_full} 

\begin{appendix}

\section{Pipeline differences} \label{app:pipeline_differences}

The JWST/NIRISS data reduction pipeline \citep{bushouse_jwst_2025} has seen continual improvements since its first release, many of which have been individually described in many works \citep[e.g.][]{pacifici_niriss_2022,taylor_niriss_2022,ravindranath_niriss_2023,plesha_improvements_2024,noirot_global_2025}.
However, to our knowledge there has not yet been any analysis which has compared the cumulative impact of these improvements on the final reduced data products.
We therefore aim to show here how these updates affect two vital outputs, the best-fit redshifts and the integrated emission line fluxes.

We use as a baseline the data reduction procedures described in detail in Section~\ref{method:data_reduction}.
All aspects of the data reduction, extraction, and analysis are kept consistent apart from those specifically mentioned below.

In particular, we compare the following versions of the data:
\begin{itemize}
    \item M22, $\rm{CRDS}=1173$: These data are almost identical to those presented in \citetalias{watson_glass-jwst_2025}, with the only difference being the addition of the diffuse background subtraction described in Section~\ref{method:data_reduction}. 
    The raw data were reduced using the JWST Operations Pipeline build 11.1, with the CRDS context set manually to 1173. 
    The wavelength calibrations for all orders used the in-flight calibrations derived in \cite{matharu_updated_2022}, which remain the default calibrations in \grizli\ as of the time of writing this manuscript.
    \item M22, $\rm{CRDS}=1413$: 
    These data were reduced using the same Operations Pipeline as in the main analysis, build 12.0.2, with fixed CRDS context 1467. The wavelength calibrations for all orders were the \cite{matharu_updated_2022} calibrations.
    \item P24: These data are the same as in the main analysis. 
    They were reduced with CRDS context 1467, and the wavelength calibrations for the first order dispersed spectra were taken from \cite{pirzkal_next_2024}. 
    All other spectral orders used the calibrations of \cite{matharu_updated_2022}.
    \item STScI: These data were again reduced using the same pipeline build, with context 1467.
    As above, the wavelength calibrations for all non-first order spectra were taken from \cite{matharu_updated_2022}.
    The first order calibrations were those used in the JWST/STScI pipeline, which were originally delivered in CRDS 1439.\footnote{Although further improvements to the wavelength calibration are expected in Operations Pipeline build 12.3, corresponding to CRDS context $\geq$1499, this was still a ``Candidate'' build as of the time of writing of this manuscript.}
\end{itemize}

\begin{figure}
    \centering
    \includegraphics[width=\columnwidth]{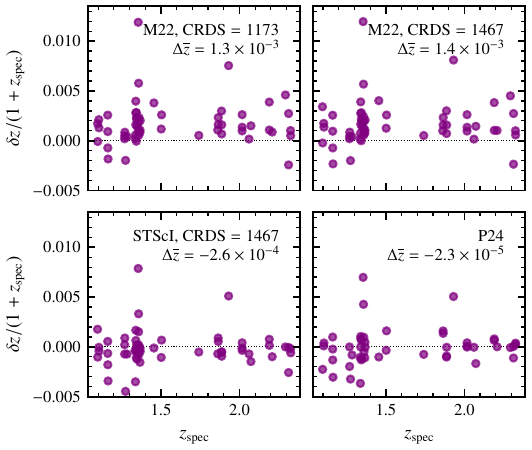}
    \caption{
    Comparison of the \grizli-derived best-fit redshifts against the spectroscopic values from the literature, for four different versions of the reduced data.
    Inset, we display the median offset of each sample.
    }
    \label{fig:redshift_comparison}
\end{figure}

For our comparison, we use the same \Halpha-selected sample as in our main analysis (see Section~\ref{method:sample_selection}), comprising 99 galaxies ranging from $1.0\leq z \leq 2.4$.
We re-extract all galaxies from each version of the reduced data, and refit the data using the default \grizli\ pipeline. 
We adopt as a central redshift estimate the values from \citetalias{watson_glass-jwst_2025}, and refit over a redshift window of $\pm10\%$.
In Fig.~\ref{fig:redshift_comparison}, we show the resulting redshift offset from other spectroscopic measurements in the literature, including \cite{mascia_glass-jwst_2024,naidu_all_2024, price_uncover_2025,bergamini_glass-jwst_2023,treu_grism_2015} (see Section~3.4 of \citetalias{watson_glass-jwst_2025} for full details of the catalogue compilation).
We use here the shorthand notation $\Delta z=\delta z/(1+\zorig)$, where $\delta z=\zorig-\zniriss$, and annotate the median offset for each reduction in Fig.~\ref{fig:redshift_comparison}.
Noting that `M22,CRDS=1173' is essentially a subset of the full catalogue presented in \citetalias{watson_glass-jwst_2025}, the observed offset here is fully consistent with that value.
The updates to the reduction pipeline by themselves appear to make little difference to the best-fit redshift, with a similar offset observed for `M22,CRDS=1413'.
However, the updated wavelength calibrations used in the `P24' and `STScI' reductions make a much larger impact.
Whilst the overall scatter is similar between all four versions ($\sigma_{\Delta z}\sim2\times10^{-3}$), the median redshift offset for `P24' and `STScI' is significantly reduced. 

\begin{figure*}
    \centering
    \includegraphics[width=\textwidth]{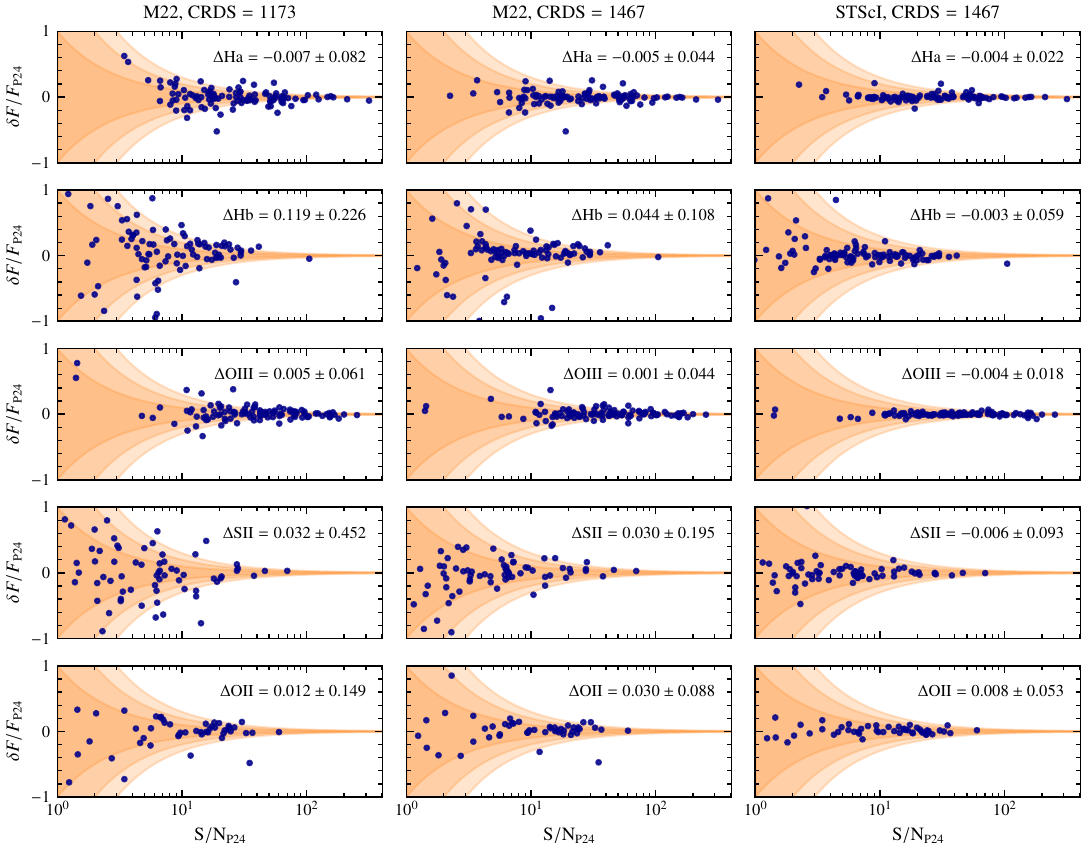}
    \caption{
    A comparison of the \grizli-derived integrated line fluxes between different versions of the reduced data.
    The data are presented as fractional offsets relative to the `P24' reduction, and plotted against the estimated \sn.
    The shaded regions correspond to the $1\sigma$, $2\sigma$, and $3\sigma$ intervals based on the \sn\ of the `P24' measurements.
    Inset, we display the median and the standard deviation of each distribution.
    }
    \label{fig:line_flux_comparison}
\end{figure*}

In Fig.~\ref{fig:line_flux_comparison}, we compare the integrated line fluxes as measured by \grizli\ for the four reductions.
We use as a baseline for our comparison the `P24' reduction used in our main analysis.
Although the sample was selected based on \Halpha\ and \Hbeta\ falling within the grism filter coverage, the wide redshift range permits us to compare the flux for several other nearby lines, namely \Sii, \Oiii\footnote{
It is crucial to note here that by default, \grizli\ fits both \Oiiiblue\ and \Oiiired\ using a single template where the line fluxes are tied in the ratio $1:2.98$.
The reported integrated \Oiii\ flux is therefore the sum of both lines, whereas the drizzled \Oiii\ emission line map contains the flux only from \Oiiired. 
}, and \Oii.

Directly comparing the two `M22' reductions, we can see the cumulative impact of improvements to the data reduction pipeline over the past year.
Across our sample, the median line flux shows only a small shift ($\approx4\%$ for \Hbeta, $\lesssim1\%$ for all other lines).
However, the scatter is considerable, and far in excess of that expected based on the estimated \sn.
Even for the highest \sn\ \Oiii\ and \Halpha\ measurements, the flux in individual galaxies can shift by $\gtrsim25\%$.
Whilst there is no clear cause for these discrepancies, the most likely candidate is the vast improvement to the WFSS sky backgrounds presented in \cite{noirot_global_2025}.
As such, we would urge caution against over-interpreting data reduced with older pipeline versions, and recommend using more recent reductions wherever possible.

Comparing between data reduced with the same pipeline version, but different wavelength calibrations, we find a much better consistency in the measurements.
The median offset for the measured lines is similar, but the scatter is greatly decreased.
Across all lines, we find that the fluxes measured using the STScI wavelength calibrations are much closer to those using the P24 calibrations, than either are to the M22 calibrations.
In particular, there are no systematic offsets observed for \Hbeta, \Sii, or \Oii, and the scatter in the measurements is generally consistent with the \sn\ of the line.
However, one remaining concern is the galaxies with the highest \sn\ emission lines.
Even limiting our search to galaxies where the $\sn\gtrsim100$ in \Halpha\ or \Oiii, we find the normalised residuals between `P24' and `STScI' to be distributed with a standard deviation $\sigma\approx4$\footnote{Although one might note that these are not truly independent measurements, but rather different attempts to perform the same measurement.}.
We therefore draw the following conclusions from this analysis:
\begin{itemize}
    \item The largest discrepancies between line flux measurements arise when comparing between data reduced with different versions of the JWST pipeline.
    \item The most recent wavelength calibrations (P24 and STScI) produce generally very similar results, but neither are fully consistent with earlier in-flight calibrations (M22).
    \item The fractional differences in the flux measured with different wavelength calibrations does not correlate with the estimated \sn\ as might be expected.
    At the highest \sn, uncertainties in the measurements from \grizli\ may be underestimated by a factor of four.
    \item The pipeline version used and the wavelength calibration adopted should always be reported for any NIRISS analysis, to allow for accurate comparisons of the measurements between different studies.
\end{itemize}

\section{Filter selection} \label{app:filter_selection}

\begin{figure*}
    \centering
    \includegraphics[width=\textwidth]{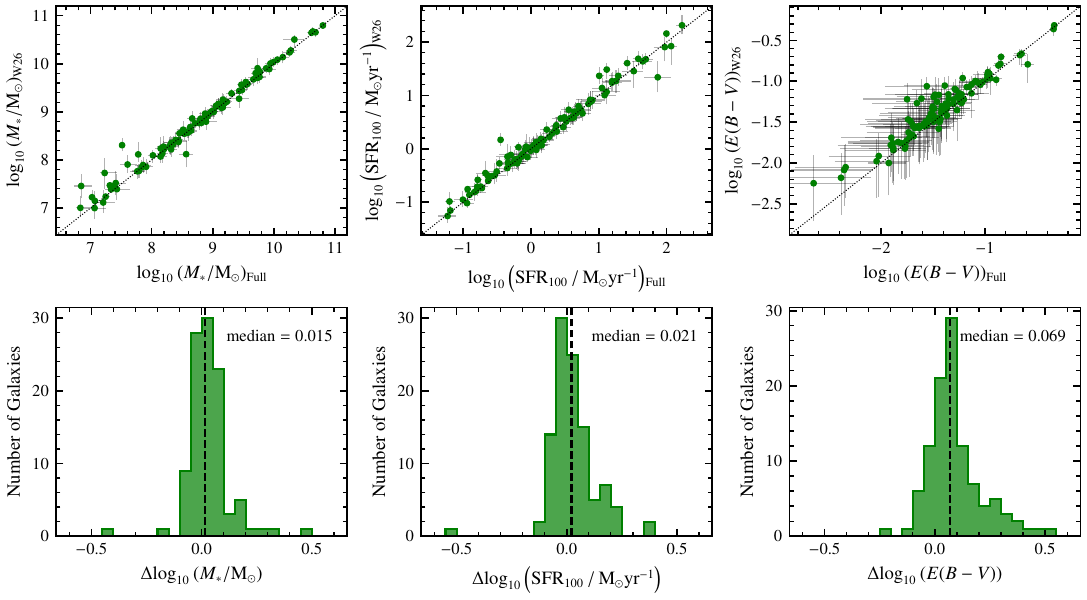}
    \caption{
    A comparison between measurements of the stellar mass, star-formation rate, and stellar reddening, derived using the full set of filters available (``Full''), and those used in the primary analysis (``W26'').
    On the bottom row, we display the histograms of the offsets, adopting the shorthand $\Delta Q\equiv Q_{\mathrm{W26}}-Q_{\mathrm{Full}}$ for quantity $Q$, with the medians of the distributions annotated.
    We show here the logarithm of the \EBV\ measurements for ease of visualisation.
    }
    \label{fig:sed_comparison_all_filters}
\end{figure*}

As we are using only a subset of the available photometry in our analysis, we assess here the impact of excluding the longer-wavelength data on our SED-derived measurements.
We make use of the MegaScience ``SUPER'' photometric catalogue \citep{suess_medium_2024}, which contains aperture-corrected fluxes measured from PSF-matched mosaics across 27 HST, JWST/NIRCam, and JWST/NIRISS wide and medium-band filters.
We cross-match this catalogue with sources in our analysis within $0\farcs3$.
Although it is not necessarily certain that the segmentation map and thus extent of all sources is consistent between this catalogue and our own analysis, this selection provides the most representative sample possible.

We use this sample to perform two sets of SED fits.
For the first, referred to as ``Full'', we use all available filters in the MegaScience catalogue, and fit the photometry using \textsc{bagpipes} with the same model components detailed in Table~\ref{tab:bagpipes_components}.
For the second, termed ``W26'', we constrain the fit to use only the photometric filters with central wavelengths $\lambda_{\textrm{c}}\lesssim2\textrm{\textmu m}$ (excluding the HST WFC3-IR filters, which have broader PSFs than the corresponding JWST filters), matching the setup used in our analysis.
We show the results for three crucial SED-derived properties in Fig.~\ref{fig:sed_comparison_all_filters}, where we take the logarithm of \EBVsed\ for ease of viewing.
For both stellar mass and SFR, we observe no systematic deviation from the 1:1 relation between the two SED fits, and $\lesssim0.02$\,dex shift in the median offset, where we adopt the shorthand $\Delta Q\equiv Q_{\mathrm{W26}}-Q_{\mathrm{Full}}$ for quantity $Q$.
This indicates that the absence of longer-wavelength data has no impact on either of these properties over the redshift range used in this analysis.

For \EBVsed, we observe a small shift ($\lesssim0.07\,$dex) towards higher attenuations when fitting over a limited wavelength range.
However, this is negligible when considering the uncertainties on either set of measurements ($\approx0.2\sigma$), and thus we do not consider it to have any significant impact on our analysis.
A full discussion of this small discrepancy in \EBVsed is beyond the scope of this analysis, but we consider it a useful avenue for future research.
In particular, we note that this shift will apply to almost all comparable datasets, including pure parallel NIRISS programmes such as PASSAGE \citep{malkan_parallel_2025}, or future Roman grism programmes such as the HLWAS \citep{observations_time_allocation_committee_roman_2025} and the Grism Reionisation and Cosmic Evolution survey (PI: Malhotra, PID 2003), and interpretations arising from these measurements should bear this limitation in mind.

\section{Merged catalogue IDs} \label{app:interacting_systems}

We tabulate here the galaxies that were designated as either interacting systems, or a single object with resolved clumpy star formation.
In both cases, we merge the previous objects (as referenced in \citetalias{watson_glass-jwst_2025}) into a single object, using the forced extraction tool \textsc{pygrife}\footnote{\href{https://github.com/PJ-Watson/pyGriFE}{github.com/PJ-Watson/pyGriFE}} to avoid recomputation of the contamination model.
The resulting sources were assigned a new ID, as shown in Table~\ref{tab:merged_ids}, 
when performing our multiregion fits.

\begin{table}[]
    \centering
    \begin{tabular}{c|c}
    \hline\hline
    New ID & IDs in \citetalias{watson_glass-jwst_2025} \\
    \hline
    10017 & 17, 19, 21 \\
    10521 & 521, 561 \\
    11689 & 1689, 1718 \\
    12250 & 2250, 2258, 2282, 2290 \\
    12355 & 2355, 2363 \\
    \end{tabular}
    \caption{The ID mapping between the old and new data reductions.}
    \label{tab:merged_ids}
\end{table}

\section{Case B template validity} \label{app:case_b_template_validity}

We assess the validity of only providing physically-motivated templates during the multiregion fitting procedure.
As our theoretical understanding of Case B violation is lacking, we do not attempt to run a full photoionisation model in order to produce physical line ratios, as might be more appropriate in regions where mechanisms such as Case A recombination holds.
Instead, we simply inject two additional template sets into the multiregion fit, whereby the emission line flux from \Hbeta\ is enhanced by 10\% (approximating the upper limit on the Balmer decrement from Case A recombination) and 25\% respectively.
As the best-fit model is a linear combination of all spectral templates, this also allows for any intermediate enhancement of \Hbeta\ relative to \Halpha.

\begin{figure}
    \centering
    \includegraphics[width=\columnwidth]{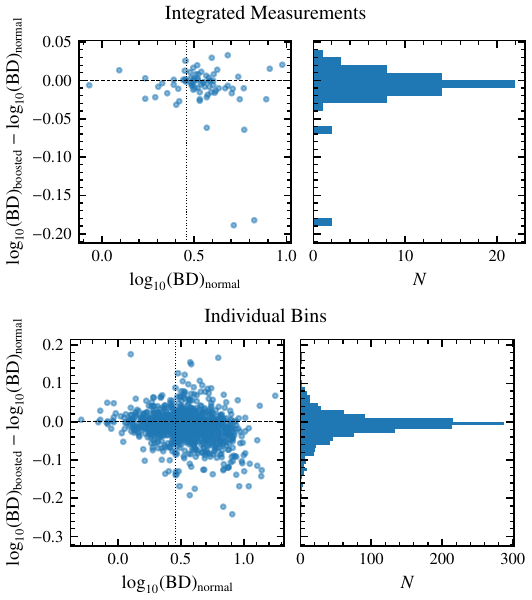}
    \caption{
    (Left) The difference in the measured Balmer decrement when fitting with \Hbeta-boosted templates over the regular \bagpipes\ fits, against the decrement measured using the regular fits.
    (Right) The histogram of the distribution.
    We show the distributions for both the integrated (top) and resolved (bottom) cases.
    }
    \label{fig:app_template_selection_comparison}
\end{figure}

We re-run the multiregion fits for all galaxies in our final sample, tripling the maximum number of sampled spectral templates in order to account for the larger parameter space.
The vast majority of our sample displayed a negligible change, and we show the difference in the measured Balmer decrement in Fig.~\ref{fig:app_template_selection_comparison} for both the integrated and resolved cases.
For the integrated measurements, we find the median offset in the Balmer decrement to be $-4\times10^{-3}$\,dex, with standard deviation $1\times10^{-2}$\,dex.
For the resolved case, the $\Hbeta$-boosted fits return lower Balmer decrements at the level of $-8\times10^{-3}$\,dex, with scatter $3\times10^{-2}$\,dex.

\begin{figure}
    \centering
    \includegraphics[width=\columnwidth]{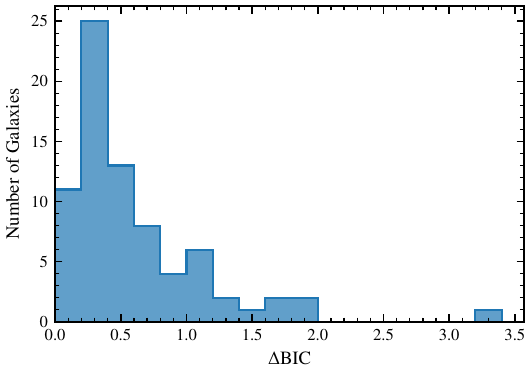}
    \caption{
    The difference in the Bayesian Information Criterion (BIC), comparing our fits with the regular \textsc{bagpipes} templates, and the \Hbeta-boosted set.
    The distribution shows a weak evidence in favour of the regular templates for all galaxies in our sample.
    }
    \label{fig:app_BIC_case_B_vs_case_A}
\end{figure}

For the majority of our sample, the total \chisq\ decreases when including the \Hbeta-boosted templates.
However, this can also be caused simply by the increased number of templates used for the modelling, i.e. a higher number of free parameters, which will inevitably improve the fit if all other variables are equal.
We assess the improvement in the fit by using the normalised Bayesian Information Criterion, formulated as:
\begin{equation}
  \mathrm{BIC} = \frac{\chi^2 + k \ln \left( n \right) }{ 2 n }.
\end{equation}
Models with a lower BIC are preferred over those with a larger BIC, and this is quantified by the difference $\Delta\mathrm{BIC}=\mathrm{BIC}_{\mathrm{high}}-\mathrm{BIC}_{\mathrm{low}}$.
We find that for all galaxies in our sample, fitting with the additional \Hbeta-boosted templates results in a higher $\mathrm{BIC}$, indicating the regular \textsc{bagpipes} templates are sufficient for fitting these data.
An alternative interpretation, perhaps more relevant to our analysis, is that the inclusion of these additional templates may lead to overfitting of the data.
We show the distribution of $\Delta\mathrm{BIC}$ in Fig.~\ref{fig:app_BIC_case_B_vs_case_A}, noting that the small values are indicative of weak statistical power supporting one model over another.

\section{NNLS optimisations} \label{app:nnls_optimisations}

\begin{figure}
    \centering
    \includegraphics[width=\columnwidth]{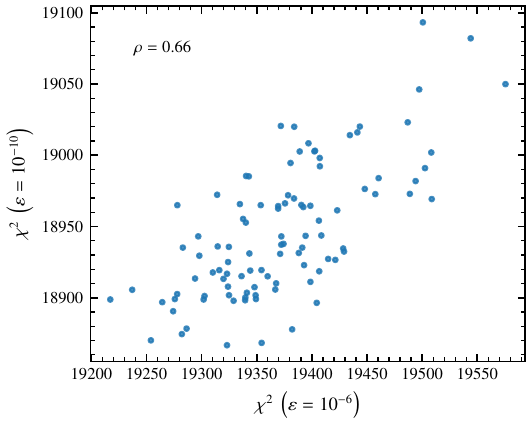}
    \caption{
    The \chisq\ statistic obtained for 100 iterations of multiregion fitting for galaxy ID 3070, with two different relative termination tolerances, denoted by $\varepsilon$.
    Inset, we display the Pearson correlation coefficient $r$.
    }
    \label{fig:NNLS_chi2}
\end{figure}

The required computation time per iteration depends on two factors -- the time needed to forward model all templates, and the time needed to find the optimum combination thereof.
The former scales linearly with the number of regions (\nregions), the number of posterior samples per region (\nsamples), and the number of pixels per beam ($m$). All three of these parameters can be tweaked to achieve an optimal but efficient sampling.
The latter is significantly more complex, and is a function of both the number of pixels per beam and the product $\nregions \times \nsamples$. 
Various non-negative least squares (NNLS) solvers have been researched and implemented since the initial algorithm proposed by \cite{lawson_solving_1974} and currently implemented in \textsc{scipy}, including Fast-NNLS \citep{bro_fast_1997} and coordinate-wise optimisations \citep{franc_sequential_2005}.
We adopt here the bounded-variable least squares (BVLS) solver described by \cite{yang_fast_2024}, and implemented in the \textsc{python} package \textsc{adelie}\footnote{\url{https://github.com/JamesYang007/adelie}}.

We display in Fig.~\ref{fig:NNLS_chi2} the results of fitting galaxy 3070 using the BVLS solver with two different tolerances for termination, denoted by $\varepsilon$, and set here to $10^{-6}$ and $10^{-10}$.
A clear correlation can be seen between the two runs, with a Spearman rank correlation coefficient of $\rho=0.66$ ($p\ll0.001$).
We note that the median times taken to calculate the best-fit template coefficients were $0.99\pm0.07$\,s for $\varepsilon=10^{-6}$, and $33.0\pm4.9$\,s for $\varepsilon=10^{10}$.
We therefore suggest that it is more efficient to focus on identifying the best input template set, and derive the optimum set of coefficients for this singular set, than to attempt to derive highly accurate coefficients for all sets of templates.

\begin{figure}
    \centering
    \includegraphics[width=\columnwidth]{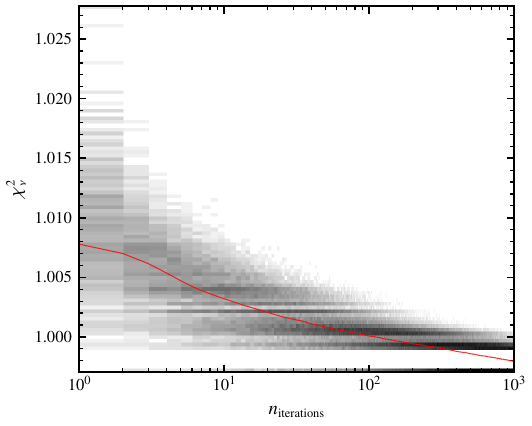}
    \caption{
    The reduced \chisq\ statistic, as a function of the number of iterations performed for the multiregion fitting. The red line denotes the mean from 1000 realisations.
    }
    \label{fig:NNLS_cumulative_chi2}
\end{figure}

With this in mind, we consider also the most efficient number of iterations to perform.
In the noise-free case, this would require sampling all combinations of the posterior distributions for all regions, a prohibitively expensive solution.
In reality, we recognise that the posteriors for each region likely include a number of degenerate solutions, and occupy a narrow region of parameter space, and thus a noise-limited solution can be achieved in far fewer iterations.
We demonstrate this empirically in Fig.~\ref{fig:NNLS_cumulative_chi2}.
We fit galaxy 3070 for 1000 iterations, using the same parameters as in the main analysis, to obtain a set of reduced \redchisq\ values.
For a number of iterations \niters, ranging between 1 and 1000, we then synthesise a fit by randomly selecting $n$ values from this set, choosing as our solution the one with the lowest \redchisq.
We perform this for 1000 realisations, to obtain a realistic distribution, and we overlay in red the mean \redchisq\ as a function of \niters.

As expected, the average \redchisq\ statistic decreases monotonically with the number of iterations performed.
Of particular interest is that even a single iteration of the multiregion fitting represents a considerable improvement over the default \grizli\ single-spectrum extraction ($\redchisq=1.73$).
Conversely, for excessively large numbers of iterations (e.g. $\niters\gtrsim10^{3}$ for this case), it becomes increasingly likely that any improvement to the \redchisq\ comes from overfitting.
This is somewhat mitigated against by the template priors for each region being physically motivated (that is, they are the posterior distributions from SED fitting), rather than an unconstrained template set as in the standard \grizli\ fitting procedure.
However, this should not be considered a panacea, and care should still be taken where the galaxy of interest contains little signal in the dispersed spectra, or is heavily contaminated.
In general, we wish to reach $\redchisq\approx1$, which is achieved here for $\niters\sim10^2$.
We therefore adopt this value throughout this analysis.

\section{Blended emission line corrections} \label{app:flux_corrections}

\begin{figure}
    \centering
    \includegraphics[width=\columnwidth]{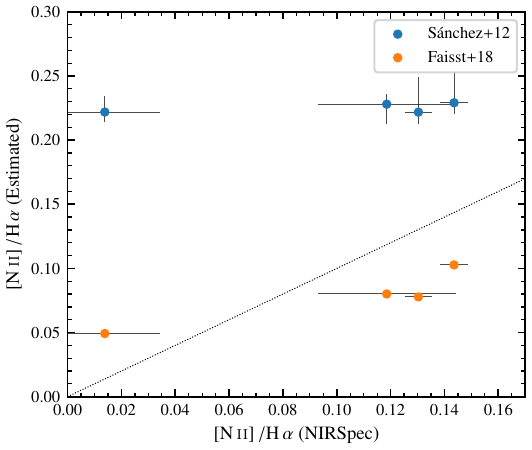}
    \caption{
    The estimated \Nii/\Halpha\ ratio via two different methods, against the direct measurement from archival JWST/NIRSpec data.
    The dotted line indicates a $1:1$ correspondence.
    }
    \label{fig:nirspec_niiha_comp}
\end{figure}

For four galaxies in our sample, archival JWST/NIRSpec observations were available from the GLASS-JWST ERS programme (PID 1324, PI T. Treu), originally reduced and presented in \cite{mascia_glass-jwst_2024}.
These utilised the high-resolution R2700 gratings (G140H, G235H, and G395H) in combination with the F100LP, F170LP, and F290LP filters, providing a spectral resolution of $R\sim1000\mathrm{-}2000$ across the 1--2.3\,\textmu m wavelength range in common with our JWST/NIRISS data.
We downloaded the fully processed spectra from the DAWN JWST Archive, with the reduction strategy described in detail in \cite{de_graaff_rubies_2025} and \cite{heintz_strong_2024}.
We used \textsc{msaexp} \citep{brammer_msaexp_2023} to fit the continuum level and measure all emission lines available using Gaussian profiles.
From these, we obtain the \Nii/\Halpha\ ratio as $(\Niiblue+\Niired)/\Halpha$, with the uncertainties propagated in quadrature.

We compare these direct measurements against the estimated \Nii/\Halpha\ ratio using two different methods.
The first is the empirical relation from \cite{faisst_empirical_2018}, which parameterises the ratio as a function of stellar mass and redshift.
We use here the total stellar mass, as measured from the sum of the spatially-resolved SED fits (see Section~\ref{method:multiregion}).
The second is the empirical relation from \cite{sanchez_integral_2012}, which links the ratio to the rest-frame dust-corrected $(B-V)$ colour.
Here, we use the spatially-resolved SED fits to derive these colour maps for the four galaxies, with the same binning scheme and quality cuts applied as detailed in Section~\ref{results:case_b:resolved}.

We show the comparison between these two relations in Fig.~\ref{fig:nirspec_niiha_comp}, as a function of the direct measurements from NIRSpec.
The uncertainties on \Nii/\Halpha\ as derived using the \cite{sanchez_integral_2012} calibration represent the full variation across all bins for each galaxy, whereas no uncertainties are available for the \cite{faisst_empirical_2018} calibration.
It is immediately clear that the \cite{sanchez_integral_2012} relation significantly overestimates the contribution of \Nii\ to the \Halpha\ emission line complex.
To some extent, this is not entirely surprising, as this relation was calibrated using more massive local galaxies, and thus may not be expected to hold for galaxies at Cosmic Noon.
Although the rest-frame colours of galaxies do evolve with redshift \citep{rudnick_rest-frame_2003}, so too does the mass-metallicity relation \citep[a][b]{sanders_mosdef_2021}, and thus we consider the relation from \cite{faisst_empirical_2018} the more accurate one to use for our analysis.

We stress that this should not be considered a fully accurate comparison, given the extremely limited sample size, and the partial coverage of the NIRSpec slit compared to the extent of the NIRISS emission (see \cite{dalmasso_quantifying_2025} for a more in-depth discussion of this latter point).
Further investigation in this area is clearly needed, in order to obtain a more accurate correction for the contribution of \Nii\ to the total \Halpha\ flux, although we defer that to future analyses.

\section{Background level} \label{app:background_levels}

\begin{figure}
    \centering
    \includegraphics[width=\columnwidth]{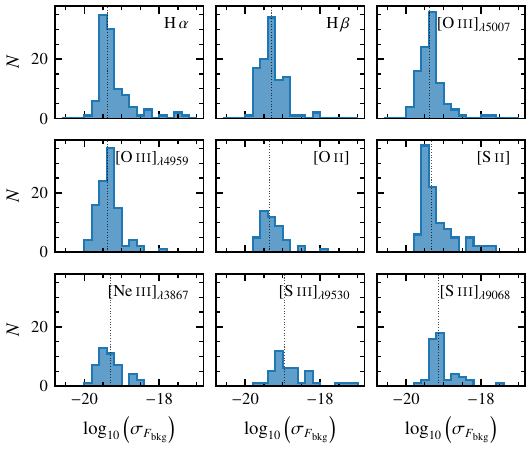}
    \caption{
    The RMS background level measured from the emission line maps of all sources in our sample.
    The dotted vertical lines indicate the median of each distribution.
    }
    \label{fig:measure_background}
\end{figure}

For each source in our sample, we attempt to measure the per-pixel background level from each of the emission line maps available.
We mask out the area covered by the segmentation map of the source itself, and the region not covered by the overlap of the dispersed beams, and measure both the median and root mean square (RMS) of the remaining values.
We find no significant deviation of the median background level from zero in our sample, justifying our assumption that the background was correctly modelled and subtracted from the dispersed grism data.
We plot the distributions of the measured background RMS in Fig.~\ref{fig:measure_background}.
The overall variation in our sample is low, with only the two \Siii\ lines showing elevated background variability, but with lower number statistics than many of the other lines.
Importantly for this analysis, there is no significant difference between the median background RMS in the \Halpha\ and \Hbeta\ emission line maps.
We therefore adopt a single value for the $1\sigma$ per-pixel background RMS, averaged over all sources and emission lines, calculated as $3.9\times10^{-20}\,\rm{erg}/\rm{s}/\rm{cm}^2/\rm{pixel}$.
Thus, considering only the background noise, the $3\sigma$ per-pixel detection limit for the GLASS-JWST ERS dataset would be $1.2\times10^{-19}\,\rm{erg}/\rm{s}/\rm{cm}^2/\rm{pixel}$.
However, as there will be additional contributions to the uncertainty owing to the continuum subtraction, and the Poisson noise of any emission line itself, we stress that this is only a lower limit to the minimum detection threshold.

\section{Binning schemes} \label{app:binning_schemes}

\begin{figure}
    \centering
    \includegraphics[width=\columnwidth]{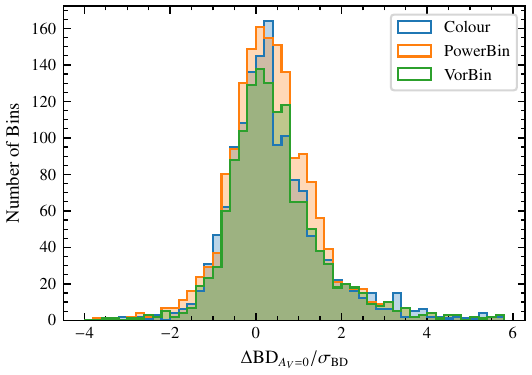}
    \caption{
    The distance of the Balmer decrement measurements from the Case B limit, normalised by their uncertainties in the same manner as in Fig.~\ref{fig:BD_distance_model_dustfree_integrated}, for each of the three binning schemes considered.
    }
    \label{fig:vorbin_powerbin}
\end{figure}

In Fig.~\ref{fig:vorbin_powerbin}, we show the distribution of resolved Balmer decrements measured using the three different binning schemes, relative to the dust-free limit assuming Case B recombination holds.
We find no significant difference in the mean or median of the distributions, with only minor changes to the proportion of galaxies more than 1$\sigma$ below the Case B limit (Colour: 5.2\%, \textsc{powerbin}: 6.5\%, \textsc{vorbin}: 5.4\%).
On conclusions regarding Case B violation are therefore not impacted by the choice of binning scheme.
However, due to the manner in which the bins are constructed, we note that both \textsc{vorbin} and \textsc{powerbin} lead to extended bins towards the edges of galaxies, even in the presence of clumpy emission.
Visually, the nearest-neighbours colour scheme better preserves the spatial information present in galaxies such as ID 2074 (see Fig.~\ref{fig:demonstrate_subtraction}), and hence, alongside the increased number of galaxies for which the binning can be performed, we prefer to use this scheme for our analysis.

\end{appendix}
\end{document}